\documentclass[preprint,12pt]{elsarticle}

\usepackage{amssymb}

\usepackage{bm}
\usepackage{color}
\usepackage{graphicx}
\usepackage{enumitem}
\usepackage{cprotect}
\usepackage{float}

\usepackage{fancyvrb}
\usepackage{upquote}
\usepackage{multirow}
\definecolor{ourcolor}{rgb}{0.7, 0.25, 0.05}

\newcommand{\qmsq}{\qmsq}
\renewcommand{\qmsq}{\q^2}

\newcommand{\be}{\begin{equation}}
\newcommand{\ee}{\end{equation}}
\newcommand{\een}{\end{subequations}}
\newcommand{\ben}{\begin{subequations}}
\newcommand{\beq}{\begin{eqalignno}}
\newcommand{\eeq}{\end{eqalignno}}

\newcommand{\lsim}{\mathrel{\mathop{\kern 0pt \rlap
      {\raise.2ex\hbox{$<$}}}\lower.9ex\hbox{\kern-.190em $ \sim$}}}
\newcommand{\gsim}{\mathrel{\mathop{\kern 0pt
      \rlap{\raise.2ex\hbox{$>$}}}\lower.9ex\hbox{\kern-.190em $\sim$}}}

\newcommand{\VectorTypefaceArrow}{
\let\oldvec\vec
\renewcommand{\vec}[1]{\oldvec{##1}} 
\newcommand{\uvec}[1]{\hat{##1}} 
}

\VectorTypefaceArrow

\usepackage{upgreek} 
\newcommand{\q}{{\widetilde{q}}}

\usepackage{mathtools}

\usepackage{booktabs}
\usepackage{xcolor}
\usepackage{arydshln}

\allowdisplaybreaks

\newcounter{bla}

\journal{Computer Physics Communications}

\begin{document}
\VerbatimFootnotes 
\begin{frontmatter}



\title{\includegraphics[width=5cm]{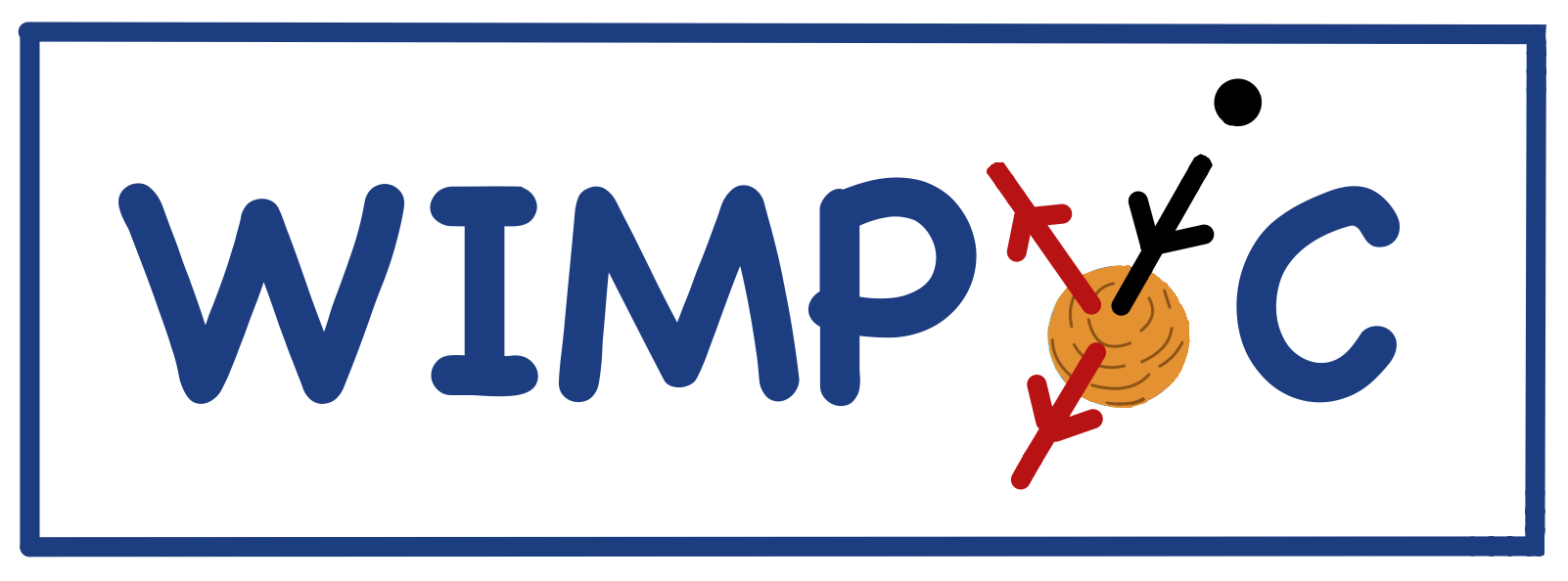}\\[0.3cm]WimPyC: an extension module of WimPyDD for the calculation of WIMP capture in celestial bodies}


\author[a,b]{Sunghyun Kang}
\author[a,b]{Stefano Scopel}
\author[c]{Gaurav Tomar}

\cortext[author] {Corresponding author.\\\textit{E-mail address:} scopel@sogang.ac.kr}

\address[a]{Department of Physics, Sogang University, Seoul 121-742, South Korea}
\address[b]{Center for Quantum Spacetime, Sogang University, 35 Baekbeom-ro, Mapo-gu, Seoul, 121-742, South Korea}
\address[c]{School of Basic Sciences and Humanities, APJ Abdul Kalam Technological University,\\ CET Campus, Thiruvananthapuram, Kerala 695016, India}

\begin{abstract}
We introduce WimPyC, a Python code for the calculation of the capture rate of Weakly Interacting Massive Particles (WIMPs) by celestial bodies through nuclear scattering in the optically thin regime. 
WimPyC is an extension of the WimPyDD code, that calculates WIMP--nucleus scattering signals in direct detection (DD) experiments, and allows to combine DD and capture in celestial bodies in virtually any scenario within the framework of Galilean--invariant non--relativistic effective theory (NREFT), including inelastic scattering, an arbitrary WIMP spin and a generic WIMP velocity distribution in the Galactic halo.  WimPyDD and WimPyC are suitable for both top--down approaches, where the interaction operators of a high--energy physics model are matched to those of the NREFT, and for bottom--up studies, where the Wilson coefficients of the NREFT are explored in a model--independent way and/or where the velocity distribution is written in terms of a superposition of streams taken as free parameters. As in the case of WimPyDD WimPyC exploits the factorization of the three main components that enter in the calculation of the capture rate: i) the Wilson coefficients that encode the dependence of the signals on the ultraviolet completion of the effective theory; ii) a response function that depends on the nuclear physics; iii) the halo function that depends on the WIMP velocity distribution. 
In WimPyC these three components are calculated and stored separately for later interpolation and combined together only as the last step of the signal evaluation procedure. This makes the phenomenological study of the capture rate with WimPyC transparent and improves computational speed.
\end{abstract}

\end{frontmatter}


{\bf PROGRAM SUMMARY}

\begin{small}
\noindent
{\em Module Title: WimPyC}  \\
{\em Licensing provisions: MIT }                                   \\
{\em Programming language: Python3}                                   \\
{\em First release: v2.0.0}                     \\
{\em Official repository: HEPForge}\\
{\em Home page: \verb|wimpydd.hepforge.org| }                                   \\

{\em Nature of problem}\\
Stars and planets in our galaxy are expected to be embedded in an extended dark matter (DM) halo made of Weakly Interacting Massive Particles (WIMPs). When they cross a celestial body DM particles can lose energy and become gravitationally captured by the same WIMP--nucleus scattering process driving direct detection in terrestrial detectors. The rate at which this process occurs is called the capture rate, and, under proper conditions, can lead to a dense population of DM particles inside the celestial body with a consequent enhancement of their annihilation rate. This can produce indirect observational signatures such as a flux of high--energy neutrinos from the celestial body, or an increase in its temperature/luminosity. Combining direct and indirect signals can potentially improve our chance to detect WIMPs, but can be cumbersome especially in generalized scenarios.\\

{\em Solution method:}\\
The WimPyC module seamlessly integrates with the WimPyDD code, that calculates WIMP--nucleus scattering signals in direct detection experiments. It allows to calculate the WIMP capture rate in virtually any scenario, including an arbitrary spin, inelastic scattering, and a non-standard halo function. Each ingredient, including the celestial body, can be set up independently in an easy and intuitive way.
\end{small}

\section{Introduction}
\label{sec:introduction}

In Ref.~\cite{WimPyDD_2022} we introduced WimPyDD, a customizable, object--oriented Python code for the calculation of expected rates of WIMP--nucleus scattering events in direct detection (DD) experiments in underground solid and liquid state detectors within the frame of the most general Non--Relativistic Effective Field Theory (NREFT) WIMP--nucleon Hamiltonian~\cite{haxton1,haxton2}. In the present paper we wish to introduce a new WimPyDD module called WimPyC, in order to extend the code to the calculation of WIMP capture in celestial bodies. Before its present release WimPyC was tested in several publications~\cite{halo_independent_sogang_eft, halo_independent_sogang_inelastic, halo_independent_sogang_long_range}.

All celestial bodies such as the Sun, the Earth, stars, planets and stellar remnants are expected to be embedded in a dark matter halo made of WIMPs. When a WIMP crosses the celestial body it can lose energy and become gravitationally captured. This can eventually lead to a dense population of WIMPs that enhances their annihilation rate, potentially producing indirect observational signatures such as a flux of high–energy neutrinos~\cite{Silk_Olive_1985, capture_gould_1987}, an increase in the temperature/luminosity of the celestial body~\cite{Goldman_Nussinov_1989}, or even to alterations in the evolutionary history of stars~\cite{dark_stars_2007, Salati_Silk_1987} that can include the formation of low-mass Black Holes incompatible with standard stellar evolution theory~\cite{capture_bh1,capture_bh2,capture_bh3,capture_bh4} or trigger runaway fusion in white dwarfs leading to Type IA supernovae~\cite{supernovae_from_capture}. In some scenarios WIMPs in celestial bodies can even lead to a gamma signal, such as in the case of annihilation through long--lived mediators~\cite{capture_jupiter_long_lived} or when in the inelastic DM scenario a large population of the heavy state  decays outside the celestial body~\cite{capture_neutron_stars_gamma}.

Since in most common scenarios DD and capture in celestial bodies are driven by the same WIMP--nucleus scattering process, combining the two signals can be useful to extend the sensitivity to the WIMP parameter space\footnote{\verb|WimPyC| does not calculate WIMP capture driven by WIMP--electron scattering~\cite{capture_electron_scattering1,capture_electron_scattering2} }. For instance, indicating with $u$ the WIMP speed in the reference frame of the solar system, due to the energy threshold DD experiments are insensitive to low values of $u$, while capture in the Sun is suppressed or impossible at high values of $u$, and only combining the two signals it is possible to obtain halo--independent bounds that, by probing the full range of incoming WIMP speeds, do not depend on the specific choice of the velocity distribution~\cite{halo_independent_single_stream, halo_independent_sogang_eft, halo_independent_sogang_inelastic, halo_independent_sogang_long_range}. Moreover, when the WIMP--nucleus interaction is described by NREFT the scattering rate is a quadratic form in the parameter space of the couplings that has flat directions corresponding to large cancellations among the Wilson coefficients of the effective Hamiltonian~\cite{sogang_lmi_2022}. Since each nuclear target has different flat directions the bounds on the Wilson coefficients can be improved by exploiting the complementarity among the targets used in DD and those that drive WIMP capture in celestial bodies. The WimPyC module handles WIMP-nucleus scattering using the same classes, routines and input data of WimPyDD, allowing to jointly analyze DD and WIMP capture in a seamless way. 

Several codes exist in the literature whose tasks partially overlap those performed by WimPyC: Asteria~\cite{asteria}, micrOMEGAs~\cite{micromegas}, DarkSUSY~\cite{darksusy}, DaMaSCUS~\cite{damascus_2}, DaMaSCUS-EarthCapture~\cite{damascus_earth}, Capt’n General~\cite{captn} and DarkCapPy~\cite{darkcappy}. All codes treat the WIMP--nucleus scattering process as elastic, and most of them handle only Spin-Independent (SI) and Spin-Dependent (SD) interactions in the Earth and the Sun, in the optically thin regime. The treatment of multiple scattering is included in Asteria for the Sun, the Earth and for Jupiter, and in DaMaSCUS only for the Earth. Capt’n General  handles NREFT interactions for a spin 1/2 particle (it can also read Solar or stellar model files and be interfaced with the MESA stellar evolution code). DarkSUSY allows to choose among different solar models and includes predefined velocity distributions, whereas micrOMEGAs is flexible in the definition of the halo function. Finally, DarkCapPy calculates capture in the Earth followed by annihilation into light mediators. With the exception of DarkSUSY and micrOMEGAs all codes assume a Maxwellian WIMP velocity distribution.
In comparison with other codes WimPyC only handles the optically thin regime. On the other hand, it allows to implement capture in virtually any celestial body and is completely flexible on the particle physics scenario and on the WIMP velocity distribution. Moreover, being an extension of WimPyDD, it is particularly suitable to study the correlation and complementarity with direct detection also with halo--independent~\cite{halo_independent_sogang_eft,halo_independent_sogang_inelastic, halo_independent_sogang_long_range} and matricial~\cite{bracketing_sogang_ibarra_2022,  sogang_lmi_2022} techniques. 

The paper is organized as follows. In Section~\ref{sec:capture} we summarize the expressions used by WimPyC to calculate the capture rate in the optically thin regime and optimized for numerical evaluation; in Section~\ref{sec:WimPyC} the components of the WimPyC module are introduced: the \verb|isotope| class in Section~\ref{sec:isotope_class}, the \verb|load_response_functions_capture| routine in Section~\ref{sec:response_functions}; the \verb|celestial_body| class in Section~\ref{sec:celestial_body_class}; in Section~\ref{sec:signal_routines} the \verb|wimp_capture|, \verb|wimp_capture_matrix| and \verb|wimp_capture_accurate| signal routines are introduced, as well as \verb|wimp_capture_geom| for the geometrical capture rate. Several explicit examples of how to use the WimPyC module, as well as the comparison of the corresponding \verb|WimPyC| output to published results are provided in Section~\ref{sec:examples}. \ref{app:initializing_celestial_body} describes in detail how to initialize a new celestial body; a description of the pre--defined \verb|celestial_body| objects included in the \verb|WimPyC| distribution is provided in \ref{app:predefined_celestial_bodies}; the \verb|wimp_capture_annihilation| routine that handles equilibrium between capture and annihilation is described in \ref{app:annihilation}. Finally, the accuracy of the interpolation method used to improve the computational speed in the calculation of the capture rate is discussed in \ref{app:accuracy}. 

\section{WIMP capture in a celestial body}
\label{sec:capture}

In the optically--thin limit the capture rate of WIMPs of mass $m_\chi$ in a celestial body of radius $R_\odot$\footnote{While $R_\odot$ is usually used to indicate the radius of the Sun, in the following we are going to use it to indicate the radius of a generic celestial body.} is given by~\cite{capture_gould_1987}: 

\begin{eqnarray}
C_{opt}\ &=&\left(\frac{\rho_\chi}{m_{\chi}}\right) \hspace{0.5mm}\int^{R_\odot}_0 dr \hspace{0.5mm} 4 \pi r^2\sum_{T} \frac{dC_{opt,T}(r)}{dV} \label{eq:cap_rate_int} \\
\frac{dC_{opt,T}(r)}{dV}\ &=&  \int du \hspace{0.5mm} f(u) 
\hspace{0.5mm} \frac{1}{u}  \hspace{0.5mm} w(u,r)^2
\label{eq:cap_rate_int_r} \nonumber\\ &&
\times~\eta_{T}(r) \hspace{0.5mm} \hspace{0.5mm} \Theta(E^\chi_{\rm max}-E^\chi_{\rm cap}) 
\int_{E_1(u,r)}^{E_2(u,r)} \frac{d\sigma_T}{dE_R}(w) dE_R
\label{eq:cap_rate}
\end{eqnarray}

\noindent with $\rho_\chi$ the density of DM at the position of the celestial body, $\eta_{T}(r)$ the number density profile for target nucleus $T$, $u$ the asymptotic speed of the incoming WIMP at large distance from the center of the celestial body, $w(u,r)^2=u^2 + v_{\rm esc}(r)^2$ the WIMP speed at the target position, $v_{\rm esc}(r)$ the escape speed at distance $r$ from its center,  and:

\begin{eqnarray}
E_1(u,r)&=& {\rm max}[E^\chi_{\rm min}(u,r),E^\chi_{\rm cap}(u)]\nonumber\\
E_2(u,r)&=& E^\chi_{\rm max}(u,r)\label{eq:e1_e2}\\
E^\chi_{\rm min, \rm max}(u,r) &=& \frac{1}{2} m_{\chi} w^2 \left[1 - \frac{\mu^2_{\chi T}}{m^2_T} 
\left(1 \pm \frac{m_T}{m_{\chi}} 
\sqrt{1 - \frac{v^2_{T*}}{w^2}}\right)^2 \right] - \delta ,
\label{eq:E_max_min}
\\
E^\chi_{\rm cap}(u) &=& \frac{1}{2} m_{\chi} u^2 - \delta .
\label{eq:E_cap}
\end{eqnarray}

\noindent In the equations above  $m_T$ is the mass of the nuclear target, $\mu_{\chi T}$ is the WIMP--nucleus reduced mass and $\delta$ the mass splitting in case of the Inelastic Dark Matter (IDM) scenario when the DM particle $\chi$ interacts with atomic nuclei exclusively by making a transition to a second state $\chi^{\prime}$ with mass $m_{\chi^{\prime}}$ = $m_\chi$ + $\delta$ endothermally ($\delta>0$)~\cite{inelastic_Tucker-Smith:2001}  or exothermally ($\delta<0$)~\cite{exothermic_dm_2010} (the elastic case is recovered when $\delta$ =0).  In particular,  $E^\chi_{\rm min, \rm max}(u,r)$ represent the minimal and maximal recoil energies of a WIMP with incoming speed $w$ while  $E^\chi_{\rm cap}(u)$ is the minimal recoil energy for capture, i.e. for the outgoing WIMP speed to be below the escape velocity after the scattering process. A peculiar feature of IDM is that WIMP--nucleus scattering has a velocity threshold given by:

\begin{equation}
v_{T^*}\equiv\left\{\begin{aligned}
\sqrt{\frac{2\delta}{\mu_{\chi T}}} ,\quad
\text{if}\;\delta> 0\\
0,\quad
\text{if}\; \delta\le 0\, .
\end{aligned}\right. 
\label{eq:vstar}
\end{equation}  

\noindent Specifically, the condition $E^\chi_{\rm max}>E^\chi_{\rm cap}$ in the integrals of Eq.~(\ref{eq:cap_rate}) ensures that $w>v_{T^*}$ and that the scattering process is kinematically possible.

Moreover:

\begin{equation}
\frac{d\sigma_T}{d E_R}(w)=\frac{2 m_T}{4\pi}\frac{c^2}{w^2}\left
     [\frac{1}{2j_{\chi}+1} \frac{1}{2j_{T}+1}\sum_{spin}|{\cal M}(q)|^2_T 
       \right ],
\label{eq:dsigma_der}
\end{equation}

\noindent is the WIMP--nucleus differential scattering cross section in terms of the scattering amplitude averaged over the initial WIMP and target spins. In the non--relativistic effective theory described by the Hamiltonian~\cite{haxton1,haxton2,krauss_spin_1,catena_krauss_spin_1,all_spins}:
\begin{eqnarray}
{\bf\mathcal{H}}(x_i,q)&=& \sum_{\tau=0,1} \sum_{j} c_j^{\tau}(x_i,q) \mathcal{O}_{j}({\bf{q}}) \, t^{\tau},
\label{eq:H}
\end{eqnarray}

\noindent the scattering amplitude is given by:

\begin{equation}
  \frac{1}{2j_{\chi}+1} \frac{1}{2j_{T}+1}\sum_{spin}|{\cal M}(q)|^2_T=
  \frac{4\pi}{2j_T+1}\sum_{\tau\tau^{\prime}}\sum_l R_l^{\tau\tau^{\prime}} (q,v)W_{l,T}^{\tau\tau^{\prime}}(q).
\label{eq:haxton_40}
\end{equation}

\noindent In Eq.~(\ref{eq:H}) we have explicitly written the
dependence of the Wilson coefficients on some parameters $x_i$ 
(which can include the WIMP mass $m_\chi$ and the mass splitting
$\delta$) arising from the matching of the Wilson coefficients of the
low--energy effective theory to its ultraviolet completion, and on the
transferred momentum $q=\sqrt{2 m_T E_R}$ (for instance through a
propagator). All $\mathcal{O}_j$ operators for $j_\chi\le$ 1/2, and some for $j_\chi$ = 1 are listed, for instance, in Table~B.1 of~\cite{WimPyDD_2022}. Moreover, in Eq.~(\ref{eq:H}), $t^0=1$, $t^1=\tau_3$
denote the $2\times2$ identity and third Pauli matrix in the nucleon
isospin space, respectively, and the isoscalar and isovector Wilson
coefficients $c^0_j$ and $c^{1}_j$ are related to those to
protons and neutrons $c^{p}_j$ and $c^{n}_j$ by
$c^{p}_j=(c^{0}_j+c^{1}_j)/2$ and $c^{n}_j=(c^{0}_j-c^{1}_j)/2$.

In Eq.~(\ref{eq:haxton_40}) the WIMP and nuclear physics are
factorized in the two $R_l^{\tau\tau^{\prime}}$ and
$W_{l,T}^{\tau\tau^{\prime}}$ functions, respectively, with
$\tau,\tau^{\prime}=0,1$ the nuclear isospin while
$l$=$M$,$\Sigma^{\prime\prime}$
,$\Sigma^{\prime}$,$\Phi^{\prime\prime}$, $\Phi^{\prime\prime}M$,
$\tilde{\Phi}^{\prime}$,$\Delta$, $\Delta\Sigma^{\prime}$ represent
one of the possible nuclear interaction types, in the single--particle
interaction limit~\cite{haxton1,haxton2}. Explicit expressions of the WIMP response functions
$R_l^{\tau\tau^{\prime}}$ are provided in~\cite{haxton2} for the set of operators ${\cal O}_j$ ($j=1,3,...,15$) valid for $j_\chi=0,1/2$ and in~\cite{all_spins} for the set of operators ${\cal O}_{X,l,s}$ for an arbitrary $j_\chi$ (for details on how WimPyDD handles the two set of interaction operators see Appendix~B of~\cite{WimPyDD_2022}).  In particular, the WIMP response functions can be decomposed as:

\begin{equation}
R_l^{\tau\tau^{\prime}}(q,w)=R_{0,l}^{\tau\tau^{\prime}}(q)+R_{1,l}^{\tau\tau^{\prime}}(q)(w^2-v_{T,min}^2),
\label{eq:vel_dep}  
\end{equation}

\noindent where:

\begin{equation}
  v_{T,min}(E_R)=\frac{1}{\sqrt{2 m_T E_R}}\left | \frac{m_TE_R}{\mu_{\chi T}}+\delta \right |,
  \label{eq:vmin}
\end{equation}

\noindent is the minimal speed an incoming WIMP needs to have in the
target reference frame to deposit energy $E_R$, with $v_{T,min}\ge v_{T*}$ ($v_{T,min}= v_{T*}$ when $E_R=E_{R*}=\mu_{\chi T}\delta/m_T$). 

As a consequence, also the cross section in Eq.~(\ref{eq:dsigma_der}) can be written as:

\begin{equation}
\frac{d\sigma_T}{d E_R}(w)=\frac{1}{w^2}\left\{ \left (\frac{d\tilde{\sigma}_T}{dE_R}\right)_0+\left (\frac{d\tilde{\sigma}_T}{dE_R}\right)_1\frac{w^2-v_{T,min}^2}{c^2}\right \}.
\label{eq:sigma_vel_dep}  
\end{equation}

As far as the nuclear response
functions $W_{l,T}^{\tau\tau^{\prime}}$ are concerned, their power / exponential expansions have been provided in
\cite{haxton2} for most targets $T$ used in DD  experiments  and in \cite{catena} for those present in typical celestial bodies such as stars, planets and white dwarfs.

If the scattering process is driven by the Hamiltonian of
Eq.~(\ref{eq:H}) the functions $(d\tilde{\sigma}_T/dE_R)_i$ appearing in Eq.~(\ref{eq:sigma_vel_dep}) are quadratic in the Wilson coefficients of the effective theory, that
can be factored out:

\begin{eqnarray}
&&\left (\frac{d\tilde{\sigma}_T}{dE_R}\right)_i (E_R)=\sum_{j,k}
c_j^{\tau}(x_i,q_0)c_{k}^{\tau^\prime}(x_i,q_0^{\prime})\left[\frac{d\tilde{\sigma}_T^i}{dE_R}\right ]_{jk}^{\tau\tau^{\prime}}(E_R),\,\,i=0,1.
\label{eq:sigma_factorization}
  \end{eqnarray}

\noindent In the expression above $q_0$, $q_0^{\prime}$ are arbitrary
momentum scales in the case when the Wilson coefficients $c_j^{\tau}$,
$c_{k}^{\tau^\prime}$ have an explicit dependence on $q$. Using in Eq.~(\ref{eq:sigma_vel_dep}) the explicit expression of the square of $v_{T,min}(E_R)$:

\begin{equation}
 v_{T,min}(E_R)^2=\frac{m_T}{2\mu_{\chi T}^2}E_R+\frac{\delta^2}{2 m_T}\frac{1}{E_R}+\frac{\delta}{\mu_{\chi T}}  
\label{eq:vmin2}
\end{equation}

\noindent one can define:

\begin{eqnarray}
    &&\left[\frac{d\tilde{\sigma}_T^{1E}}{dE_R}\right ]_{jk}^{\tau\tau^{\prime}}(E_R)\equiv E_R \left[\frac{d\tilde{\sigma}_T^{1}}{dE_R}\right ]_{jk}^{\tau\tau^{\prime}}(E_R)\nonumber\\
    &&\left[\frac{d\tilde{\sigma}_T^{1E^{-1}}}{dE_R}\right ]_{jk}^{\tau\tau^{\prime}}(E_R)\equiv\frac{1}{E_R} \left[\frac{d\tilde{\sigma}_T^{1}}{dE_R}\right ]_{jk}^{\tau\tau^{\prime}}(E_R),
    \label{eq:diff_sigma1}
\end{eqnarray}

\noindent leading to the following set of four differential response functions:

\begin{equation}
    \left[\frac{d\tilde{\sigma}_T^a}{dE_R}\right ]_{jk}^{\tau\tau^{\prime}}=\left \{ \left[\frac{d\tilde{\sigma}_T^0}{dE_R}\right ]_{jk}^{\tau\tau^{\prime}}, \left[\frac{d\tilde{\sigma}_T^1}{dE_R}\right ]_{jk}^{\tau\tau^{\prime}},\left[\frac{d\tilde{\sigma}_T^{1E}}{dE_R}\right ]_{jk}^{\tau\tau^{\prime}},\left[\frac{d\tilde{\sigma}_T^{1E^{-1}}}{dE_R}\right ]_{jk}^{\tau\tau^{\prime}}\right \}
    \label{eq:diff_sigma}
\end{equation}

\noindent and the corresponding integrated response functions:

\begin{eqnarray}
  &&\left [\bar{\sigma}^a_T\right ]_{jk}^{\tau\tau^{\prime}}(E_R)\equiv \int_0^{E_R} dE^{\prime}_R \left[\frac{d\tilde{\sigma}_T^a}{dE_R}\right ]_{jk}^{\tau\tau^{\prime}}(E_R^{\prime}),\,\,\,a=0,1,1E, 1E^{-1}. \nonumber\\
  \label{eq:sigma_tilde}
\end{eqnarray}

\noindent Using Eqs.~(\ref{eq:sigma_vel_dep}) and (\ref{eq:sigma_factorization}) in (\ref{eq:cap_rate}) the capture rate can then be written as:

\begin{eqnarray}
  &&\frac{dC_{opt,T}(r)}{dV}=\sum_{k=1}^{N_s} \delta\eta_k\times \frac{d C_{opt,T}}{dV}(r,u_k)\nonumber\\
  && \frac{d C_{opt,T}}{dV}(r,u_k)\equiv \eta_{T}(r) \sum_{jk}\sum_{\tau\tau^{\prime}} c_j^{\tau}(x_i,q_0)c_{k}^{\tau^\prime}(x_i,q_0^{\prime})  \nonumber\\
  && \Theta \left(E_2(u_k,r)-E^\chi_{\rm cap}(u_k)\right) \times\nonumber\\
  &&  \left \{\left[\bar{{\sigma}}^0_{T}\right ]_{jk}^{\tau\tau^{\prime}}(E_R)
  +\left(\frac{u^2_k+v^2_{esc}(r)}{c^2}-\frac{\delta}{\mu_{\chi T}}\right)\left [
  \bar{{\sigma}}^1_{T}\right ]_{jk}^{\tau\tau^{\prime}}(E_R) \right .\nonumber\\
&& \left .  -\frac{m_T}{2\mu_{\chi T}^2}
  \left [\bar{{\sigma}}^{1E}_{T}\right ]_{jk}^{\tau\tau^{\prime}}(E_R)-\frac{\delta^2}{2 m_T}
  \left[\bar{{\sigma}}^{1E^{-1}}_{T}\right ]_{jk}^{\tau\tau^{\prime}}(E_R)\right \}^{E_2(u_k,r)}_{E_1(u_k,r)},\nonumber \\
 && {} \label{eq:capture_master}
\end{eqnarray}

\noindent with $\{f(E_R)\}_{E_1}^{E_2} \equiv f(E_2)-f(E_1)$. Moreover, in Eq.~(\ref{eq:capture_master}) the WIMP speed distribution\footnote{Eqs.~(\ref{eq:cap_rate_int_r}) was derived in~\cite{capture_gould_1987}, where it was pointed out that the calculation of the total capture rate includes a sum over all possible incoming WIMP directions, effectively averaging over the angular distribution.} $f(v)\equiv 1/(4\pi) \int d\Omega v^2f(\vec{v})$ is written as a superposition of streams:

\begin{equation}
f(u)=\sum_k^{N_s} \lambda_k \delta(u-u_k),
\label{eq:f_streams}
\end{equation}

\noindent so that the halo-function is written as:

\begin{eqnarray}
 \eta(u)&=&\int_{u}^{\infty}\frac{f(u^{\prime})}{u^{\prime}}\,du^{\prime}=\sum_{k=1}^{N_s}
  \delta\eta_k\theta(u_k-u),\nonumber\\
  \delta\eta_k&=&\lambda_k/u_k.
\label{eq:eta}  
\end{eqnarray}

Combined with the radial integration of Eq.~(\ref{eq:cap_rate_int}), Eq.~(\ref{eq:capture_master}) serves as the master formula used by WimPyC to calculate the capture rate through Eq.~(\ref{eq:cap_rate_int_r}). It has the same structure of Eq.~(20) of Ref.~\cite{WimPyDD_2022} used by WimPyDD to calculate expected rates for DD, and it exploits explicitly the factorization of the Wilson coefficients $c_j^{\tau}$, the
response functions $\left [\bar{\sigma}^a_T\right ]_{jk}^{\tau\tau^{\prime}}(E_R)$ ($a$=0, 1, 1$E$, 1$E^{-1}$) and of the halo functions $\delta\eta_k^{(0,1)}$ (see Appendix E of~\cite{WimPyDD_2022} for how to implement the latter in WimPyDD). Since the response functions $\left [\bar{\sigma}^a_T\right ]_{jk}^{\tau\tau^{\prime}}(E_R)$ depend only on the single external argument $E_R$ WimPyC speeds
up the calculation by tabulating each $\left [\bar{\sigma}^a_T\right ]_{jk}^{\tau\tau^{\prime}}(E_R)$ for later
interpolation. In particular such tables do not depend on the
WIMP mass $m_\chi$ and on the mass splitting $\delta$, that only enter in the mapping between the $u_k$ values and the
energies $E_1(u_k)$ and $E_2(u_k)$ at which the response
functions in the tables are interpolated. We have extensively verified that such interpolation method leads to accurate results, although, for completeness and validation purposes \verb|WimPyC| also provides an alternative (and slower) routine that performs the full integrations of Eq.~(\ref{eq:cap_rate}). 

In analogy to DD rates, thanks to the factorization in Eq.~(\ref{eq:sigma_factorization}) the capture rate can be seen as a quadratic form in the couplings:

\begin{eqnarray}\nonumber
    &&C_{opt}=c^T \cdot {\cal M}\cdot c,\\
    && c=\{ c_j^\tau \}
    \label{eq:cap_rate_matrix}.
\end{eqnarray}
\noindent This allows to apply matricial techniques when analyzing the multi--dimensional parameter space of the WIMP effective Hamiltonian~\cite{brenner_2020, bracketing_sogang_ibarra_2022, sogang_lmi_2022}. \verb|WimPyC| includes a dedicated routine for the calculation of the matrix ${\cal M}$ (see example in Section~\ref{subsec:matrix}).

Moreover, again in the same way of DD, using $\delta \eta_k^{(0,1)}=\lambda_k^{(0,1)}/v_k$ Eqs.~(\ref{eq:cap_rate_int_r},\ref{eq:capture_master}) can be recast in the form:

\begin{eqnarray}
&&C_{opt}=\sum_{k=1}^{N_s} \lambda_k {\cal H}_C (u_k)\label{eq:Hk},\\
 &&{\cal H}_C (u_k)=\left(\frac{\rho_\chi}{m_{\chi}}\right) \hspace{0.5mm}\int^{R_\odot}_0 dr \hspace{0.5mm} 4 \pi r^2\sum_{T} \frac{1}{u_k}\frac{dC_{opt,T}(r,u_k)}{dV},
 \label{eq:lambda_k}
\end{eqnarray} 

\noindent which corresponds to the expression:

\begin{equation}
C_{opt}=\int_{0}^\infty d u {\cal H}_C(u) f(u), 
\label{eq:HC}
\end{equation}
\noindent 

 \noindent when the velocity distribution $f(u)$ is parameterized in terms of a sum of streams using Eq.~(\ref{eq:f_streams}). Eq.~(\ref{eq:HC}) shows that, by making use of the master formula of Eq.~(\ref{eq:capture_master}),  
 \verb|WimPyC| is very suitable to apply the halo-independent method discussed in~\cite{gondolo_eta2, generalized_halo_indep, Kahlhoefer_halo_independent}, where the velocity distribution or the halo function is written in terms of a set of free parameters (namely, the velocities $v_k$ and either the $\lambda_k$'s or the $\eta_k$'s, see the discussion in Section 2.1 of~\cite{WimPyDD_2022}). This will be illustrated in the example of  Section~\ref{sec:halo_independent}.

Eqs.~(\ref{eq:cap_rate}, \ref{eq:capture_master}) assume that a WIMP is captured for good after a single scattering event off a nuclear target at distance $r$ from the center of the celestial body brings the WIMP below the escape velocity $v_{esc}(r)$. Specifically, after the first scattering the WIMP is assumed to be locked into a bound orbit so that over the age of the system it keeps crossing the celestial body scattering time and again, eventually thermalizing with the nuclei in its environment and settling into a distribution peaked at its center.\footnote{\verb|WimPyC| does not handle situations where the thermalization process can be more involved (for instance in the case of inelastic scattering~\cite{capture_neutron_stars_gamma, thermalization_inelastic_Blennow_2018}).} 

Thanks to gravitational acceleration capture allows to probe the WIMP--nucleus scattering process at WIMP speeds not within the reach of terrestrial detectors, potentially improving the sensitivity to the WIMP parameter space. This is especially true in specific scenarios, such as inelastic scattering~\cite{sogang_wd_2022}.  However, Eq.~(\ref{eq:capture_master}) is based on the NREFT Hamiltonian of Eq.~(\ref{eq:H}) and on non--relativistic kinematics, so it is not valid if the WIMP speed becomes relativistic inside the celestial body. In particular, this happens for neutron stars, for which WimPyC cannot be used. 

The optically thin approximation of Eq.~(\ref{eq:cap_rate}) is valid as long as the WIMP-nucleus cross section is small enough to neglect multiple scattering. A discussion of such effect is beyond the scope of the \verb|WimPyC| code (for an accurate treatment see~\cite{asteria,multiple_scattering_Bramante,multiple_scattering_Dasgupta, multiple_scattering_Ilie}). In the limit of very large cross sections the multi--scatter regime is eventually saturated by a geometrical capture rate independent of the WIMP-nucleus interaction, when every WIMP crossing the celestial body is captured, given by:
\begin{equation}
C_{geom} = \pi R^2 \left( \frac{\rho_{\chi}}{m_{\chi}} \right) \int_0^{\infty} 
du \hspace{0.5mm} \frac{f(u)}{u} \hspace{0.5mm} w^2(R).
\label{eq:C_geom}
\end{equation}
\noindent
A reasonable approximation of the capture rate beyond the optically thin regime can be obtained by interpolating between the optically thin limit of Eq.~(\ref{eq:cap_rate}) and the geometrical capture rate, i.e. by using $C=min(C_{opt}, C_{geom})$ (see, for instance, the bottom plot of Fig.~\ref{fig:jupiter_wds_ds}). So a rule-of-thumb estimate of the validity of the optically thin regime and of the reliability of the \verb|WimPyC| routines is that $C_{opt} \ll C_{geom}$. The \verb|WimPyC| module includes a simple routine to calculate $C_{geom}$ (see Section~\ref{sec:signal_routines}).

\section{The WimPyC module}
\label{sec:WimPyC}

The WimPyC extension adds to WimPyDD the new classes and functions necessary for the calculation of WIMP capture in celestial bodies within the formalism discussed in Section~\ref{sec:capture}. It is included in the latest version of the WimPyDD distribution, that can be downloaded from: 

\begin{Verbatim}[frame=single,xleftmargin=0.5cm,xrightmargin=0.5cm,commandchars=\\\{\}]
  https://wimpydd.hepforge.org
\end{Verbatim}

\noindent The present introduction will mostly
describe default features, with a few mentions to possible
generalizations. Use the \verb|help()| function from the Python command line to get additional information about any component of the code not included here. 

\begin{figure}
\begin{center}
  \includegraphics[width=0.9\columnwidth]{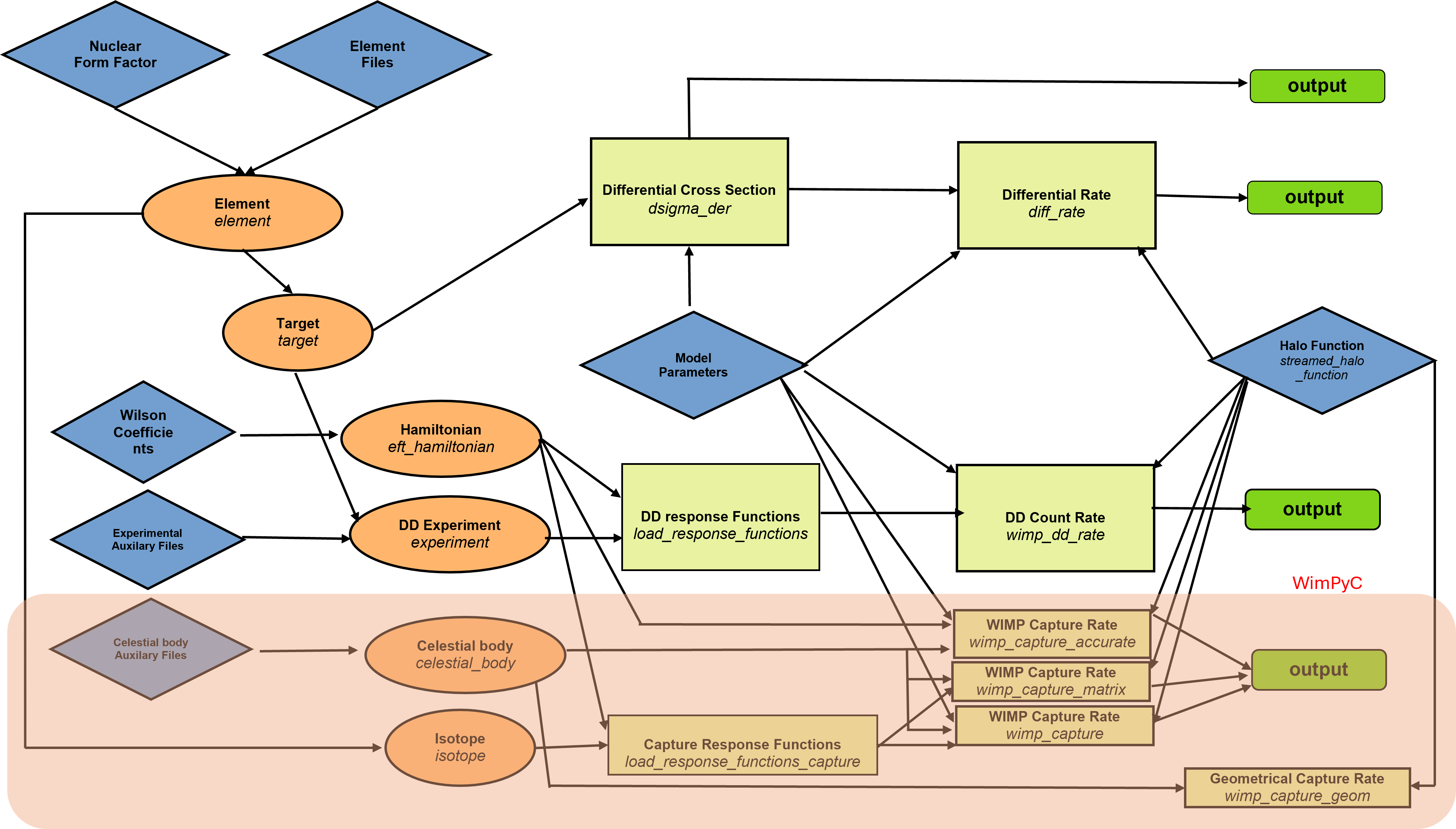}
  \end{center}
\caption{
Main combined structure of WimPyDD and WimPyC (the shaded area includes the main components of the new module). Diamond--shaped elements
indicate the input from the user, oval shapes contain classes and
rectangles represent functions. \label{fig:code_structure}}
\end{figure}

The structure of the code is summarized in
Fig.~\ref{fig:code_structure}. In such figure the shaded area contains the new classes and routines added by \verb|WimPyC|, with diamond--shaped elements indicating the input from the user, classes represented by oval shapes and functions by rectangles. In particular, the calculation of $C_{opt}$ requires to initialize a celestial body object through the class \verb|celestial_body|, and makes use of a new class \verb|isotope|, directly derived from the \verb|element|
class already existing in WimPyDD. The halo function information used in the calculation of $C_{opt}$ or $C_{geom}$ is the same used by WimPyDD to calculate DD signals.

\subsection{Installation}
\label{sec:installation}

To install the latest version of WimPyDD, including WimPyC, type:

\begin{Verbatim}[frame=single,xleftmargin=0.5cm,xrightmargin=0.5cm,commandchars=\\\{\}]
git clone https://repo.hepforge.org/source/WimPyDD.git
\end{Verbatim}

\noindent in the working directory, or download the code from
WimPyDD's homepage. This will create the folder WimPyDD, that, starting from version 2.0.0, includes a WimPyC subfolder.  All the
libraries required by the code are standard, such as matplotlib,
numpy, scipy and pickle. A joint import of the functions and classes of both WimPyDD and WimPyC is obtained by typing something like:

\begin{Verbatim}[frame=single,xleftmargin=0.5cm,xrightmargin=0.5cm,commandchars=\\\{\}]
import WimPyDD as WD
\end{Verbatim}

\subsection{New routines}
\label{sec:signal_routines}
We introduce in this Section the new routines of the WimPyC module. They work in the same way of those already existing in WimPyDD, taking some of the same inputs and assuming the same default values for them. In particular, in this and in the following sections we will only describe the features not already present in the original WimPyDD distribution, and address the reader to Ref.~\cite{WimPyDD_2022} for the others.  

To calculate the WIMP capture rate $C_{opt}$ in the optically thin limit (in s$^{-1}$) of Eq.~(\ref{eq:cap_rate}) with the expression of Eq.~(\ref{eq:capture_master}) use:

\begin{Verbatim}[frame=single,xleftmargin=0.5cm,xrightmargin=0.5cm,commandchars=\\\{\}]
WD.wimp_capture(celestial_body_obj, hamiltonian, vmin,
delta_eta, mchi,**args)
\end{Verbatim}

\noindent with \verb|celestial_body_obj| an object that defines the celestial body (see Section~\ref{sec:celestial_body_class}),  \verb|mchi| the WIMP mass $m_\chi$ in GeV, \verb|vmin| and \verb|delta_eta| two arrays in km/s and (km/s)$^{-1}$, respectively, containing the $u_k$ and $\delta\eta_k$  quantities for the halo function defined in Eq.~(\ref{eq:eta})  (WimPyDD provides the routine \verb|streamed_halo_function| to calculate them, see Appendix E of Ref.~\cite{WimPyDD_2022}) and \verb|hamiltonian| defines the effective Hamiltonian of Eq.~(\ref{eq:H}) (initialized following the instructions provided in Section 3.3.3 of~\cite{WimPyDD_2022}). If present, the arguments $x_i$ of the Wilson coefficients $c_j^\tau$ are passed by \verb|**args|. For all the other inputs and their default values (such as the WIMP local density $\rho_\chi$, the mass splitting $\delta$, etc.) check the Python prompt help (when applicable, they are the same as, for instance, in the \verb|wimp_dd_rate| routine for direct detection).  The matrix ${\cal M}$ of Eq.~(\ref{eq:cap_rate_matrix}) can be obtained by using:

\begin{Verbatim}[frame=single,xleftmargin=0.5cm,xrightmargin=0.5cm,commandchars=\\\{\}]
M=WD.wimp_capture_matrix(celestial_body_obj, hamiltonian,
vmin, delta_eta, mchi, **args)
\end{Verbatim}

\noindent The output matrix \verb|M| returned by \verb|wimp_capture_matrix| is in the isospin basis $c=\{c_j^\tau\}$.
A dictionary containing the correspondence between the matrix indices and the couplings can be obtained using:

\begin{Verbatim}[frame=single,xleftmargin=0.5cm,xrightmargin=0.5cm,commandchars=\\\{\}]
WD.get_mapping(hamiltonian)
\end{Verbatim}

\noindent The matrix \verb|M| can be rotated into a new matrix \verb|M_pn| in the proton-neutron basis $c=\{c_j^{(p,n)}\}$ using

\begin{Verbatim}[frame=single,xleftmargin=0.5cm,xrightmargin=0.5cm,commandchars=\\\{\}]
U_pn=WD.rotation_from_isospin_to_pn(hamiltonian)
M_pn=np.dot(U_pn.T,np.dot(M,U_pn))
\end{Verbatim}

A dictionary with the mapping between the indices of \verb|M_pn| and the couplings $c_i^{(p,n)}$ is obtained using:

\begin{Verbatim}[frame=single,xleftmargin=0.5cm,xrightmargin=0.5cm,commandchars=\\\{\}]
WD.get_mapping(hamiltonian, pn=True)
\end{Verbatim}

In order to speed-up the calculations the \verb|wimp_capture| and \verb|wimp_captur|
\verb|e_matrix| routines interpolate the response functions defined in Eq.~(\ref{eq:sigma_tilde}). This approach drastically reduces the computational time compared to a full integration, and we have verified its accuracy at the percent level in a very extensive class of cases (for different celestial bodies, momentum and velocity suppressed interactions, inelastic scattering, massless mediator). However, for completeness and validation purposes the \verb|WimPyC| distribution includes the routine \verb|WD.wimp_capture_accurate| that performs the calculation without approximations. In particular, for each value of the distance $r$ from the center of the celestial body and of the asymptotic velocity $u$ such routine obtains a single function of the recoil energy performing all the sums appearing in Eq.~(\ref{eq:capture_master}) (i.e. summing over the couplings combinations, the nuclear isospin and the index $a = 0, 1, 1E, 1E^{-1}$) {\it before} calculating explicitly the energy integrals from $E_1(u,r)$ to $E_2(u,r)$. By avoiding interpolations and by inverting energy integrations and linear combinations this procedure allows to verify the level of accuracy of the \verb|wimp_capture| and \verb|wimp_capture_matrix| routines, also in the case of large cancellations among the different terms in the sums of Eq.~(\ref{eq:capture_master}). To use it, type:

\begin{Verbatim}[frame=single,xleftmargin=0.5cm,xrightmargin=0.5cm,commandchars=\\\{\}]
WD.wimp_capture_accurate(celestial_body_obj, hamiltonian, 
vmin, delta_eta, mchi,**args)
\end{Verbatim}

\noindent The \verb|WD.wimp_capture_accurate| routine is significantly slower than \verb|wimp_ca|
\verb|pture| and is supposed to be used for validating purposes. One can limit its scope using arguments such as $\verb|targets_list|$, \verb|tau_range|, \verb|tau_prime_range| and \verb|n_vel_list| to include in the sums only some targets, some values of $\tau$ and $\tau^{\prime}$ and of the index $a$ of Eq.~(\ref{eq:sigma_tilde}).   

Finally, the geometrical capture rate of Eq.~(\ref{eq:C_geom}) can be calculated using:

\begin{Verbatim}[frame=single,xleftmargin=0.5cm,xrightmargin=0.5cm,commandchars=\\\{\}]
WD.wimp_capture_geom(celestial_body_obj, mchi, vmin, 
delta_eta, rho_loc=0.3)
\end{Verbatim}

An example of the output of the \verb|wimp_capture_geom| routine is represented by the horizontal line in the bottom plot of Fig.~\ref{fig:jupiter_wds_ds}.

\subsection{New classes}
\label{sec:classes}

\subsubsection{Celestial body}
\label{sec:celestial_body_class}
{\bf Main attributes:} \verb|mass| (mass in solar masses), \verb|name| (string with name), \verb|r_vec| (array with radius sampling, normalized to the interval $0-1$), \verb|rho_tot| (array with the total density radial profile $\rho_{tot}(r)$ tabulated over \verb|r_vec|), \verb|radius| (radius in solar radii), \verb|mass_fractions| (dictionary of total mass fractions $f_i$ of different targets), \verb|rho_i| (dictionary returning for each target the density profile $\rho_i(r)$ normalized to the total mass fraction, \verb|target_names| (list of strings containing the names of the targets in the celestial body - corresponds to the keys of the dictionaries \verb|mass_fractions|  and \verb|rho_i| ),  \verb|targets| (list of \verb|isotope| objects defining the targets, \verb|v_esc| (array with the escape velocity as a function of the radius) and \verb|T_c| (core temperature in Kelvin, used by the \verb|wimp_capture_annihilation| routine, see \ref{app:annihilation}).

The input of the class is a string with the name of the celestial body that corresponds to a folder containing the required information:

\begin{Verbatim}[frame=single,xleftmargin=0.5cm,xrightmargin=0.5cm,commandchars=\\\{\}]
sun=WD.celestial_body('Sun')\\
\end{Verbatim}

In the example above all the information about the Sun is loaded from the folder \verb|WimPyDD/WimPyC/Celestial_bodies/Sun| (details on how to initialize such directory are provided in \ref{app:initializing_celestial_body}).

For instance:

\begin{Verbatim}[frame=single,xleftmargin=0.5cm,xrightmargin=0.5cm,commandchars=\\\{\}]
integrate.simps(4*np.pi*sun.r_vec**2*sun.rho_tot, 
sun.r_vec)\\
1.00\\
sun.mass\\
1.0\\
integrate.simps(4*np.pi*earth.r_vec**2*earth.rho_tot, 
earth.r_vec)\\
1.00\\
integrate.simps(4*np.pi*sun.r_vec**2*sun.rho_i['4He'], 
sun.r_vec)\\
0.301927022320300217\\
sun.mass_fractions['4He']\\
0.301927022320300217\\
earth=WD.celestial_body('Earth')\\
earth.mass\\
3.002732577e-06\\
integrate.simps(4*np.pi*earth.r_vec**2*earth.rho_i['28Si']
, earth.r_vec)\\
0.14456334097116127\\
earth.mass_fractions['28Si']\\
0.14456334097116127\\
\end{Verbatim}

\subsubsection{Isotope}
\label{sec:isotope_class}

\begin{Verbatim}[frame=single,xleftmargin=0.5cm,xrightmargin=0.5cm,commandchars=\\\{\}]
WD.isotope(symbol, element=None, verbose=True)
\end{Verbatim}

{\bf Main attributes:} \verb|symbol| (string with formula of isotope), \verb|element| (optional, \verb|WimPyDD| \verb|element| object containing the isotope information), \verb|mass| (isotope mass in GeV), \verb|spin| (isotope spin), \verb|z| (atomic number), \verb|a| (mass number), \verb|func_w| (nuclear form factors), \verb|diff_response_functions_capture| 
and \verb|response_functions_capture|(differential and integrated response functions for the calculation of the WIMP capture rate).

At instantiation the \verb|isotope| class extracts the information about the isotope from an \verb|element| object that contains it. In particular, if the \verb|element| argument is not passed the class initializes it using the information contained in the folder \verb|WimPyDD/Targets|. For instance, with the following input:

\begin{Verbatim}[frame=single,xleftmargin=0.5cm,xrightmargin=0.5cm,commandchars=\\\{\}]
si28=WD.isotope('28Si')
\end{Verbatim}

\noindent the \verb|isotope| class strips the string \verb|'28Si'| of the leading digits, initializes internally an element object issuing the command \verb|element_object=WD.elemen|
\verb|t('Si')| (that is successful if the file \verb|Si.tab| is present in the folder \verb|WimPyDD/|
\verb|Targets|) and looks for the string \verb|'28Si'| in the \verb|element_object.isotopes| attribute in order to extract the relevant information.
The nuclear response functions necessary for the calculation of the WIMP--nucleus scattering process and taken from~\cite{haxton2,catena} are available in WimPyDD for 22 default elements (see~\cite{WimPyDD_2022}) that can be listed issuing the command:

\begin{Verbatim}[frame=single,xleftmargin=0.5cm,xrightmargin=0.5cm,commandchars=\\\{\}]
WD.list_elements()
\end{Verbatim}

So, in order to add a new isotope it is necessary to initialize first an element containing it in the \verb|WimPyDD/Targets| directory. The procedure to add new elements with the corresponding nuclear form factors, or to use alternative nuclear form factors for existing targets is provided in Appendix D of Ref.~\cite{WimPyDD_2022}. \verb|WimPyC| comes with a set of pre--defined isotopes that can be listed by issuing:

\begin{Verbatim}[frame=single,xleftmargin=0.5cm,xrightmargin=0.5cm,commandchars=\\\{\}]
WD.list_isotopes()
Available isotope objects in WimPyDD:
import WimPyDD as WD
WD.C12 WD.C13 WD.N14 WD.N15 WD.O16 WD.O17 WD.O18 WD.H1 
WD.Ne20 WD.Na23 WD.Mg24 WD.Al27 WD.Si28 WD.P31 WD.S32 
WD.He3 WD.Ar40 WD.Ca40 WD.He4 WD.Fe56 WD.Ni58
Use print() to get info on each isotope. For instance:
Type print(WD.Fe56)
symbol 56Fe, atomic number 26, mass 52.136
Nuclear form factor:Definition from Phys.Rev.C 89 (2014) 
6, 065501 (e-Print: 1308.6288[hep-ph]) used
for nuclear W functions as default
\end{Verbatim}

\subsection{Response functions for capture}
\label{sec:response_functions}

Given the effective Hamiltonian \verb|eft_hamiltonian_obj| and the target $T$ = \verb|isotope_obj| the differential response functions $\left[\frac{d\tilde{\sigma}_T^a}{dE_R}\right ]_{jk}^{\tau\tau^{\prime}}(E_R)$ defined in Eq.~(\ref{eq:diff_sigma}) and the tabulated values of the functions $\left [\bar{\sigma}^a_T\right ]_{jk}^{\tau\tau^{\prime}}(E_R)$  ($a = 0, 1, 1E, 1E^{-1}$) of Eq.~(\ref{eq:sigma_tilde}) are loaded by the routine:

\begin{Verbatim}[frame=single,xleftmargin=0.5cm,xrightmargin=0.5cm,commandchars=\\\{\}]
WD.load_response_functions_capture(input_obj, hamiltonian,\\
j_chi=0.5)
\end{Verbatim}

\noindent where the WIMP spin \verb|j_chi| is set to 1/2 by default. \verb|input_obj| can either belong to the \verb|isotope| class or to the \verb|celestial_body| class (in the latter case the response functions are loaded for all the targets in the celestial body).
The differential and integrated response functions are loaded in the \verb|diff_response_functions_capture| and \verb|response_functions_capture| attributes of \verb|isotope| objects, which are empty dictionaries when the isotope is initialized.  In particular, the instructions:  

\begin{Verbatim}[frame=single,xleftmargin=0.5cm,xrightmargin=0.5cm,commandchars=\\\{\}]
c12_15=WD.eft_hamiltonian('c12_15',\{12: lambda: [1,1],\\ 
15: lambda: [1,1]\})\\
Al27=WD.isotope('27Al')\\
j_chi=0.5\\
WD.load_response_functions_capture(Al27,c12_15,j_chi=0.5)\\
tau=0\\
tau_prime=0\\
c12_15.coeff_squared_list\\
[(12, 12), (12, 15), (15, 12), (15, 15)]\\
ij=1\\
ci,cj=c12_15.coeff_squared_list[ij] \\
n_vel=2 # 0,1,1E, 1Em1
dsigma_der=Al27.diff_response_functions_capture['c12_15',
j_chi, ci, cj][tau, tau_prime, n_vel]\\
E_R,sigma_bar=Al27.response_functions_capture['c12_15', 
j_chi][ij][n_vel][tau][tau_prime]
\end{Verbatim}

\noindent initialize an effective Hamiltonian $c_{12}{\cal O}_{12}+c_{15}{\cal O}_{15}$, the differential response function $\left[\frac{d\tilde{\sigma}_{^{27}Al}^{1E}}{dE_R}\right ]_{12,15}^{\tau\tau^{\prime}}(E_R)$, and the two arrays \verb|E_R| and \verb|sigma_bar| that tabulate the integrated response function $\left [\bar{\sigma}^{1E}_{^{27}Al}\right ]_{12,15}^{00}(E_R)$ as a function of the recoil energy. In both cases the $(i,j)=(12,15)$ coupling combination is set by the index \verb|ij=1| of one of the entries of the \\ \verb|eft_hamiltonian_obj.coeff_squared_list| attribute, while \verb|n_vel=2| identifies $1E$ in the list $(0,1,1E,1E^{-1})$\footnote{The functions in the \verb|diff_response_functions_capture| attribute return the $\left[\frac{d\tilde{\sigma}_T^a}{dE_R}\right ]_{jk}^{\tau\tau^{\prime}}$ functions in units of [cm$^2$ km$^2$ s$^{-2}$ keV$^{-1}$, cm$^2$ km$^2$ s$^{-2}$ keV$^{-1}$, cm$^2$ km$^2$ s$^{-2}$ GeV keV$^{-1}$, cm$^2$ km$^2$ s$^{-2}$ GeV$^{-1}$ keV$^{-1}$] for $a = 0, 1, 1E, 1E^{-1}$. The tabulated values of the $\left [\bar{\sigma}^a_T\right ]_{jk}^{\tau\tau^{\prime}}$ are in the units of the $\left[\frac{d\tilde{\sigma}_T^a}{dE_R}\right ]_{jk}^{\tau\tau^{\prime}}$ times keV. }.

In the example above the \verb|E_R| and \verb|sigma_bar| tabulated values are read from the file \verb|27Al_c_12_c_15.npy| contained in \verb|WimPyDD/WimPyC/Response|
\verb|_functions/spin_1_2/| folder. If such file does not exist it is created and the integrals of Eq.~(\ref{eq:sigma_tilde}) are calculated and saved in it.  The convention for the file names is identical to the DD case, see Section 3.4 of~\cite{WimPyDD_2022} (``Handling the response functions set") for a detailed discussion. In particular, a different file is saved for each of the coupling combinations listed in the \verb|coeff_squared_list| attribute of the \verb|eft_hamiltonian| object.  
Notice that the $\bar{\sigma}$ functions are attributes of the \verb|isotope| objects, and not of the \verb|celestial_body| objects. This implies that 
the same file is read for a given isotope, irrespective on the celestial body where it is contained. 

The distribution of \verb|WimPyC| includes a set a of tabulated functions that reproduce all the examples listed in Section~\ref{sec:examples} and on the code website \verb|wimpydd.hepforge.org|. By default the routine tabulates $\left [\bar{\sigma}^a_T\right ]_{jk}^{\tau\tau^{\prime}}(E_R)$ for 1000 values of the integration upper bound $E_R$ (the number of sampled points and the minimal and maximal values of $E_R$ can be changed using the arguments \verb|n_sampling| and \verb|er_min_cut|, \verb|er_max_cut|). 

The response functions provided by default in the \verb|WimPyC| distribution have been extensively tested to verify that their sampling is accurate enough in a very wide range of cases, including inelastic scattering, a massless propagator, interference among different interaction operators, etc. Nevertheless, the parameter space of WIMP capture is particularly extended, with very different kinematic regimes across different celestial bodies, nuclear targets and interactions types. As a consequence, in \ref{app:accuracy} we provide a discussion of this aspect, showing how the user can easily check possible under--sampling issues that may arise in particular corners of the parameter space.  In case of need, the sampling of the response functions can be increased by passing to the \verb|load_response_functions_capture| routine an array of energy values \verb|er_sampling| in keV. Moreover, setting \verb|increase_sampling=|
\verb|True| overwrites existing tabulated functions including the additional points, while \verb|force_recalculation=True| recalculates them from scratch. The array \verb|er_sampling| can be provided independently by the user or obtained for the object \verb|target| belonging to the \verb|isotope| class using:

\begin{Verbatim}[frame=single,xleftmargin=0.5cm,xrightmargin=0.5cm,commandchars=\\\{\}]
er_sampling=
WD.get_wimp_capture_response_functions_energy_sampling(
target, celestial_body_list, mchi_list, delta_list, vmin, 
n_sampling=1000)
\end{Verbatim}

\noindent where \verb|celestial_body_list| contains a list of \verb| celestial_body| objects, \verb|mchi_list| contains a list of $m_\chi$ values, \verb|delta_list| a list of $\delta$ values, \verb|vmin| an array with values of the WIMP asymptotic speed $u$, and \verb|n_sampling| fixes the length of the returned array, whose energy values are tuned on the input parameters by calculating a sample of the $E_1$, $E_2$ energies of Eq.~(\ref{eq:e1_e2}). 

\section{Examples of capture rate calculations}
\label{sec:examples}

In the following, we provide a few examples for the calculation of the WIMP capture rate. They use implementations of the celestial bodies that are included in the distribution of \verb|WimPyC|. To get their list type:

\begin{Verbatim}[frame=single,xleftmargin=0.5cm,xrightmargin=0.5cm,commandchars=\\\{\}]
WD.list_celestial_bodies()
WD.Sun WD.Earth WD.Jupiter WD.White_Dwarf WD.MS_Star\\
\end{Verbatim}

\noindent See \ref{app:predefined_celestial_bodies} for some details about their implementation. 

\subsection{Sun, NREFT operators}
\label{subsec:sun}
\begin{figure}[h!]
    \centering
    \includegraphics[width=0.45\textwidth]{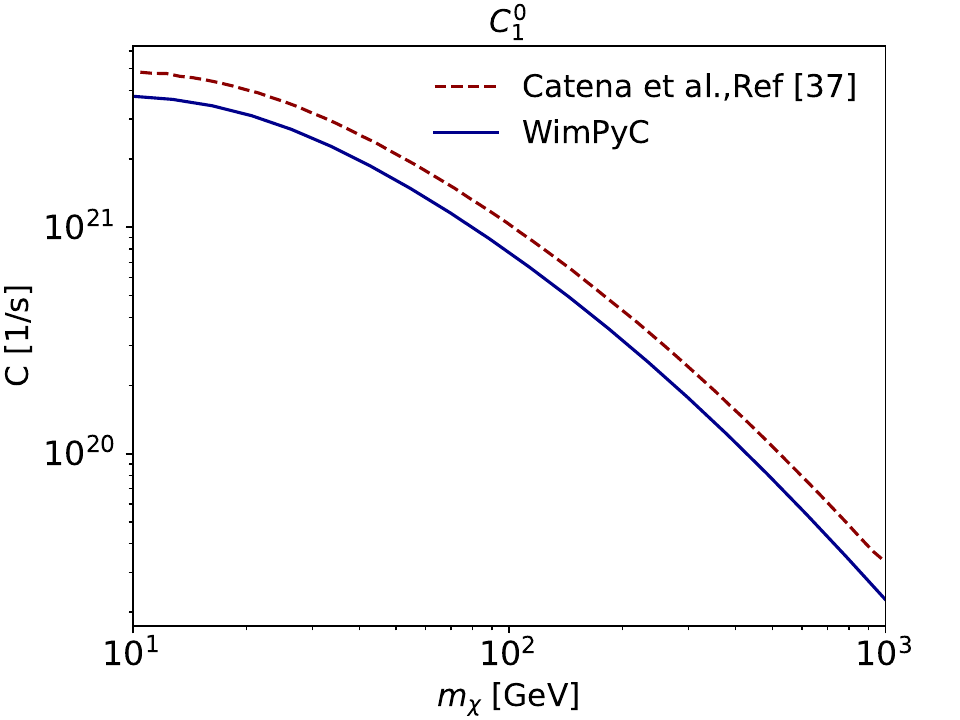} \hspace{-0.3cm}
    \includegraphics[width=0.45\textwidth]{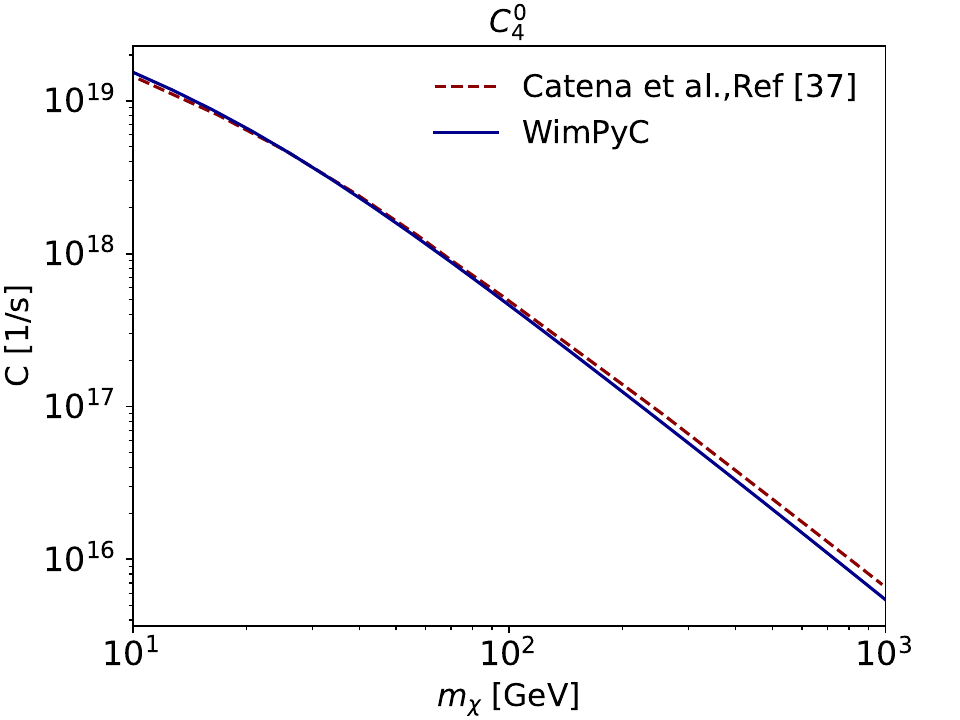} \\[0.2cm]
    \includegraphics[width=0.45\textwidth]{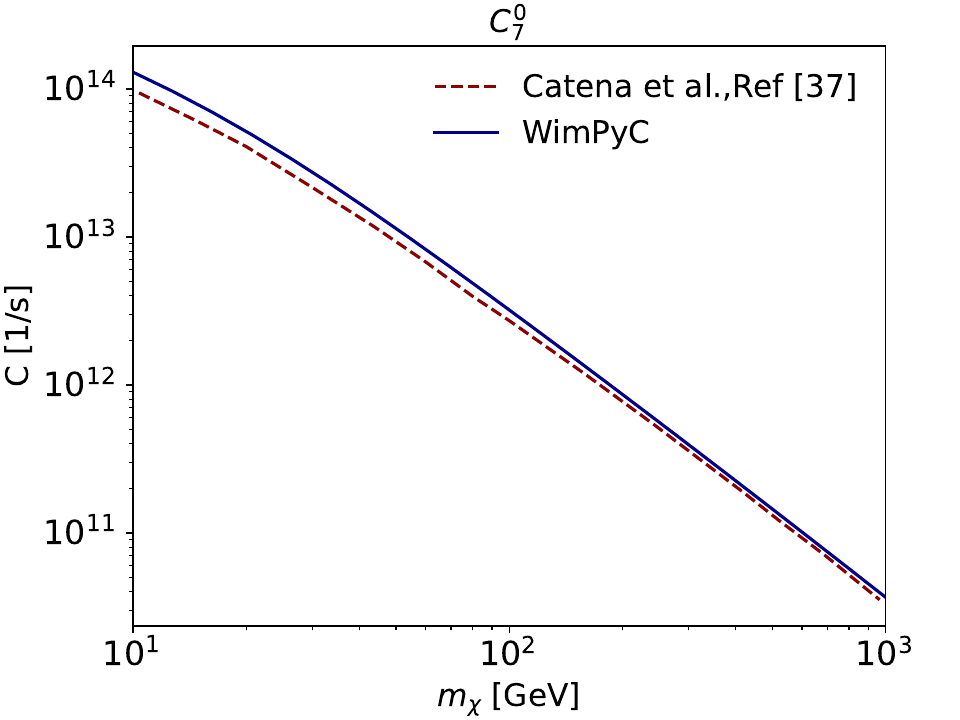} \hspace{-0.3cm}
    \includegraphics[width=0.45\textwidth]{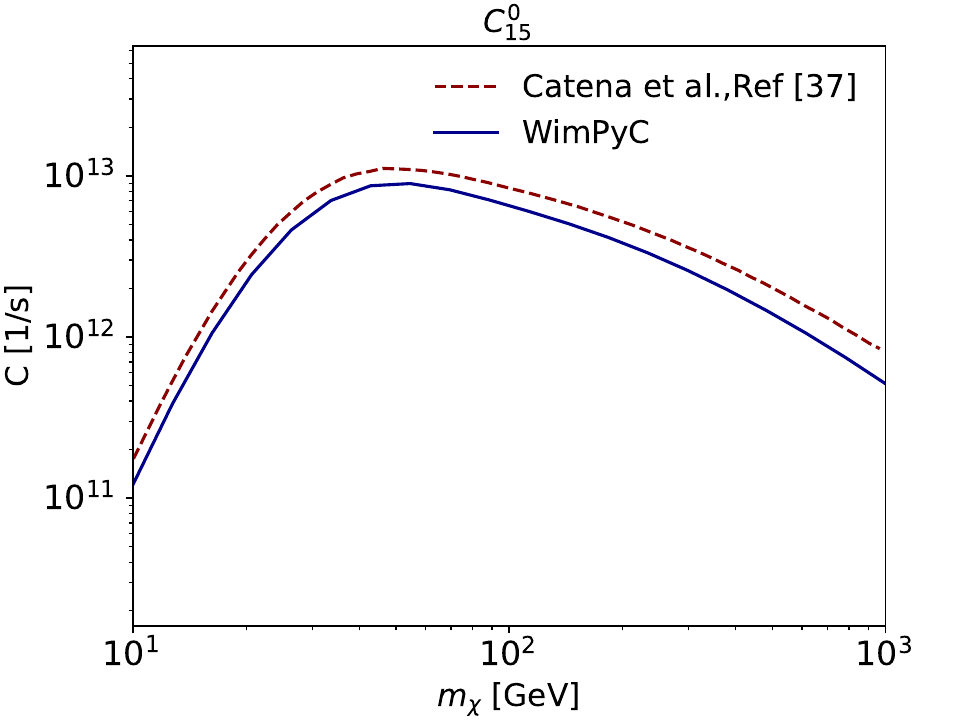}
    \caption{The WimPyC capture rate for the Sun, using the selected isoscalar NREFT operators, is compared with the results of Ref.~\cite{catena}.}
    \label{fig:nreft_sun}
\end{figure}

In the example below the capture rate in the Sun is calculated for the ${\cal O}_1$, ${\cal O}_4$, ${\cal O}_7$ and ${\cal O}_{15}$ operators of the effective Hamiltonian of Eq.~(\ref{eq:H}). In particular, following Ref.~\cite{catena} in all cases an isoscalar interaction with $c^0=10^{-3}/m_v^2$, $m_v$ = 246.2 GeV, $c^1$ = 0 is adopted, as well as a standard Maxwellian velocity distribution with $\rho_\chi = 0.4$ GeV/cm$^3$. In Fig.~\ref{fig:nreft_sun} the output of WimPyC is compared to the corresponding results of Ref.~\cite{catena}.

\begin{Verbatim}[frame=single,xleftmargin=0.5cm,xrightmargin=0.5cm,commandchars=\\\{\}]
vmin,delta_eta=WD.streamed_halo_function(v_esc_gal=550) 
mchi_vec=np.logspace(1, 3, 50)
### repeat for NREFT operators (1, 4, 7, 15)
c7 = WD.eft_hamiltonian('c7', \{7 : lambda c0, c1: 
np.array([c0, c1])\})
mv=246.2
capture=[WD.wimp_capture(WD.Sun, c7, vmin, delta_eta, 
mchi, rho_loc=0.4, c0=1e-3/mv**2, c1=0) for mchi in 
mchi_vec]
\end{Verbatim}

\subsection{Sun, SI interaction, inelastic scattering}
\begin{figure}[h!]
    \centering
    \includegraphics[width=0.6\textwidth]{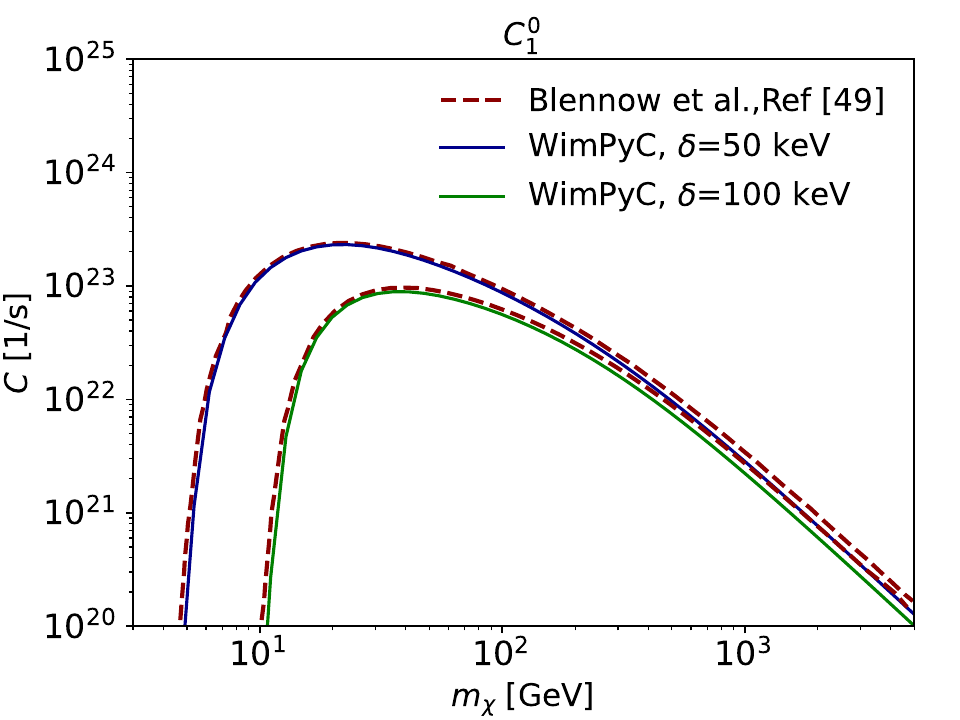}
    \caption{The WimPyC capture rate for the Sun in the inelastic scenario is compared with published results~\cite{Blennow:2015hzp}, considering $\sigma_{\chi{\cal N}}=10^{-42}$ cm$^2$ for a spin-independent isoscalar interaction.}
    \label{fig:inelastic_sun}
\end{figure}

In this example we compare the output of WimPyC with that of Ref.~\cite{Blennow:2015hzp} for WIMP capture in the Sun driven by inelastic scattering. Following Ref.~\cite{Blennow:2015hzp} we adopt an isoscalar SI interaction (${\cal O}_1$ effective operator) parameterized in terms of the WIMP--nucleon cross section, $\sigma_{\chi{\cal N}}$ = $(c^{\cal N})^2 \mu_{\chi {\cal N}}^2/\pi$, with ${\cal N}$ = proton, neutron, $1/\mu_{\chi, {\cal N}}= 1/m_\chi+1/m_{\cal N}$ using the function:

\begin{Verbatim}[frame=single,xleftmargin=0.5cm,xrightmargin=0.5cm,commandchars=\\\{\}]
WD.c_tau_SI(sigma_p, mchi, cn_over_cp=1)
\end{Verbatim}

\noindent with \verb|cn_over_cp| the ratio $c_1^n/c_1^p$ between the WIMP--neutron and the WIMP--proton couplings (see Ref.~\cite{WimPyDD_2022} and online help). The output for inelastic scattering is obtained passing to \verb|WD.wimp_capture| a non--vanishing value of the \verb|delta| parameter in keV.  

In Fig.~\ref{fig:inelastic_sun} the output of WimPyC for $\sigma_{\chi {\cal N}}$ = $10^{-42}$cm$^2$, and $\delta$ = 50, 100 keV is compared to the corresponding one of Ref.~\cite{Blennow:2015hzp} for an isoscalar interaction (for details of how to tune the halo function using the \verb|WD.streamed_halo_fun|
\verb|ction| routine see Appendix E of~\cite{WimPyDD_2022}).
\begin{Verbatim}[frame=single,xleftmargin=0.5cm,xrightmargin=0.5cm,commandchars=\\\{\}]
vmin,delta_eta=WD.streamed_halo_function(v_esc_gal=544) \\
mchi_vec=np.logspace(np.log10(4), np.log10(5000), 50)
SI=WD.eft_hamiltonian('SI',\{1: WD.c_tau_SI\})\\
### repeat for delta=50, 100
delta=50
capture=[WD.wimp_capture(WD.Sun,SI,vmin,delta_eta,mchi,
delta=delta,rho_loc=0.4,sigma_p=1e-42) for mchi in mchi_vec]
\end{Verbatim}

\subsection{Earth, NREFT operators}
\begin{figure}[h!]
    \centering
    \includegraphics[width=0.45\textwidth]{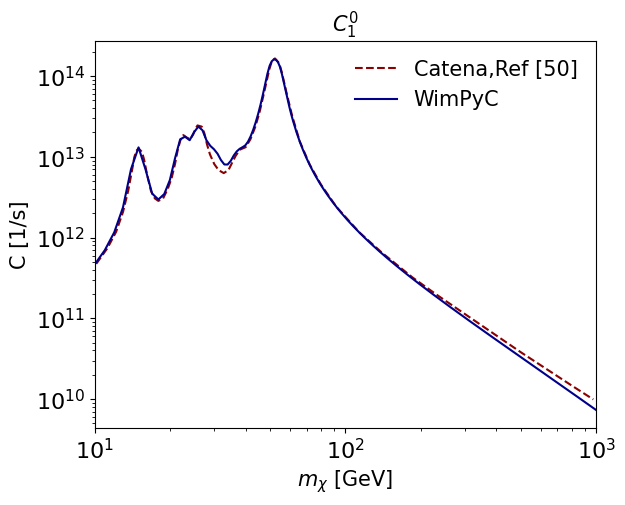} \hspace{-0.3cm}
    \includegraphics[width=0.45\textwidth]{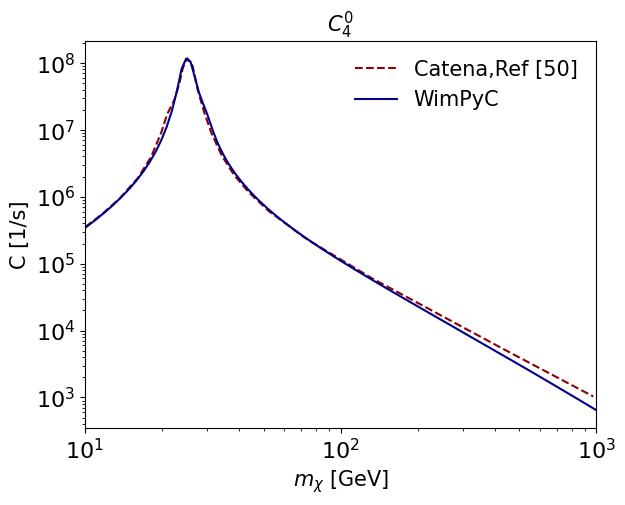} \\[0.2cm]
    \includegraphics[width=0.45\textwidth]{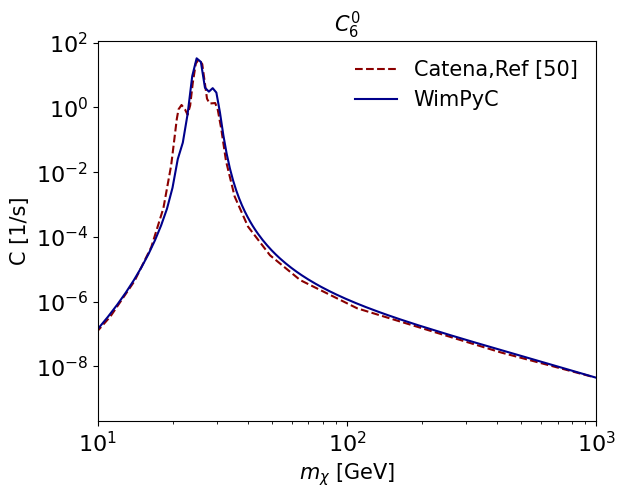} \hspace{-0.3cm}
    \includegraphics[width=0.45\textwidth]{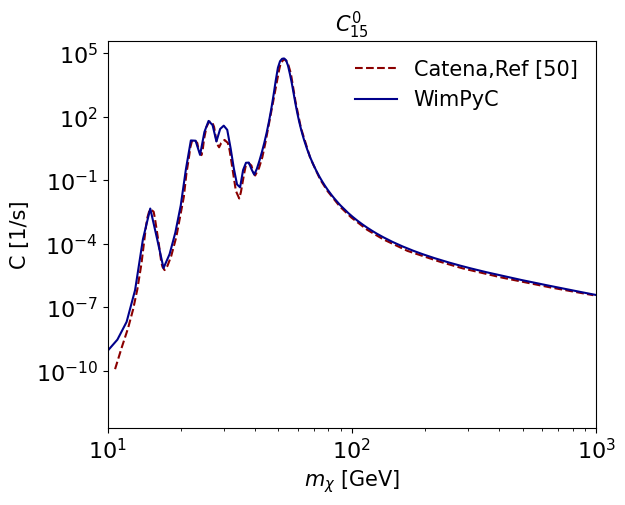}
    \caption{The WimPyC capture rate for Earth, using the selected isoscalar NREFT operators, is compared to the results of Ref.~\cite{catena_earth}.}
    \label{fig:nreft_earth}
\end{figure}

This example is analogous to that of Section~\ref{subsec:sun}. In particular, following Ref.~\cite{catena_earth} we adopt $c^0=10^{-3}/m_v^2$, $m_v$ = 246.2 GeV, $c^1$ = 0 as well as a standard Maxwellian velocity distribution with $\rho_\chi = 0.4$ GeV/cm$^3$. In Fig.~\ref{fig:nreft_earth}  the WimPyC output is compared to that of Ref.~\cite{catena_earth} for the NREFT operators ${\cal O}_i$, $i=1,4,6,15$.

\begin{Verbatim}[frame=single,xleftmargin=0.5cm,xrightmargin=0.5cm,commandchars=\\\{\}]
vmin,delta_eta=WD.streamed_halo_function(v_esc_gal=533) \\
mchi_vec=np.logspace(1, 3, 50)
### repeat for NREFT operators (1, 4, 6, 15)
c6 = WD.eft_hamiltonian('c6',
\{ 6: lambda c0, c1: np.array([c0, c1])\})\\
mv=246.2
capture=[WD.wimp_capture(WD.Earth,c6,vmin,delta_eta,mchi, 
rho_loc=0.4, c0=1e-3/mv**2, c1=0) for mchi in mchi_vec]
\end{Verbatim}

\subsection{Jupiter, SD interaction}
\begin{figure}[h!]
    \centering
    \includegraphics[width=0.45\textwidth]{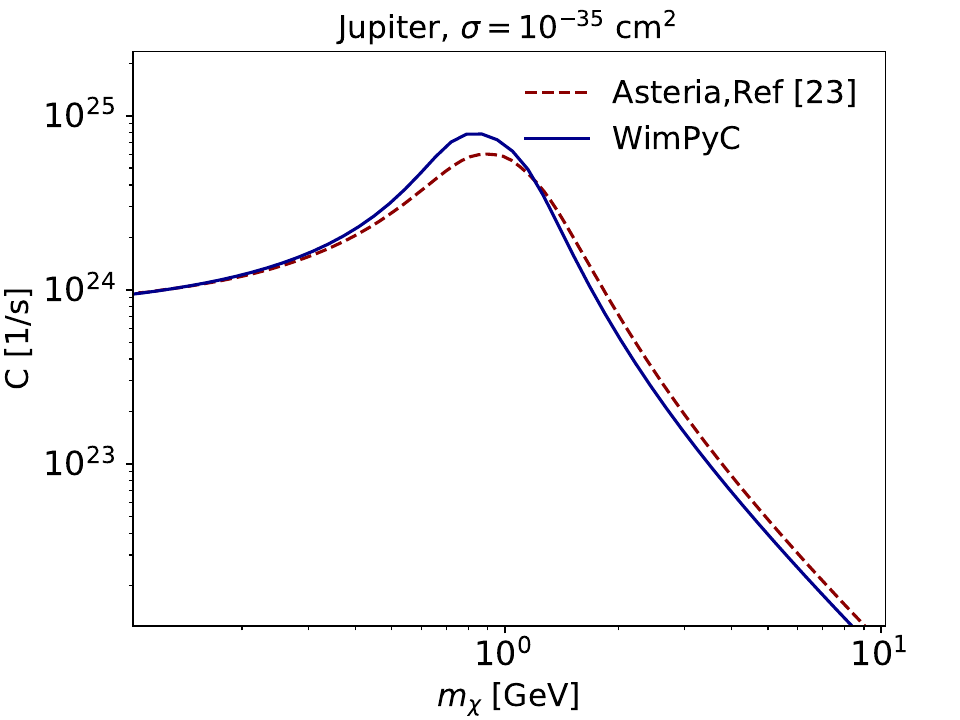} \hspace{-0.3cm}
    \includegraphics[width=0.45\textwidth]{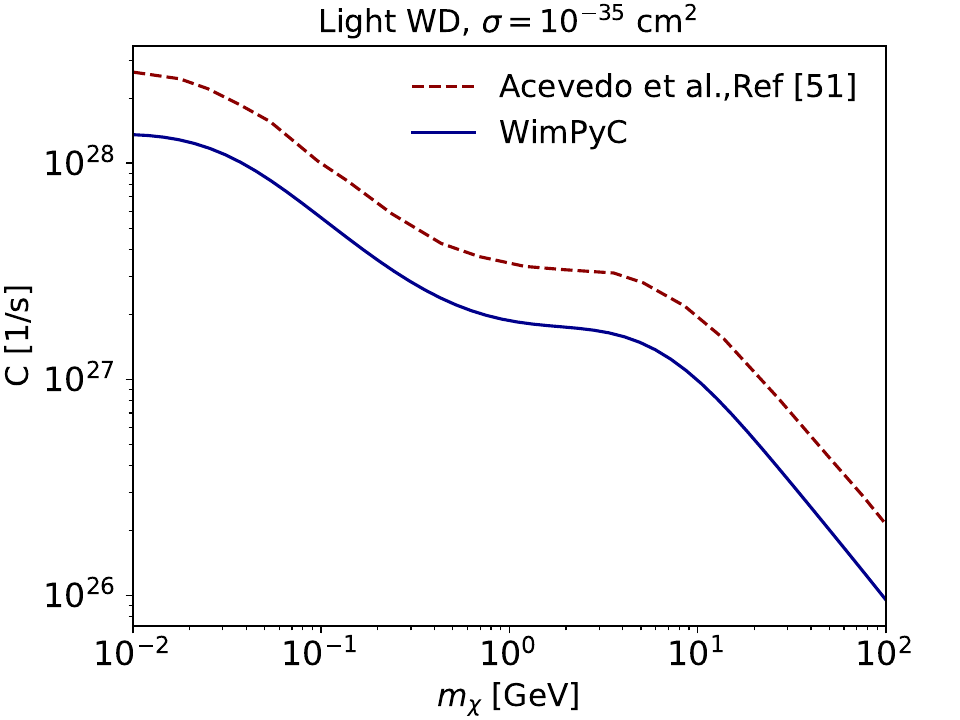} \\[0.2cm]
        \includegraphics[width=0.5\textwidth]{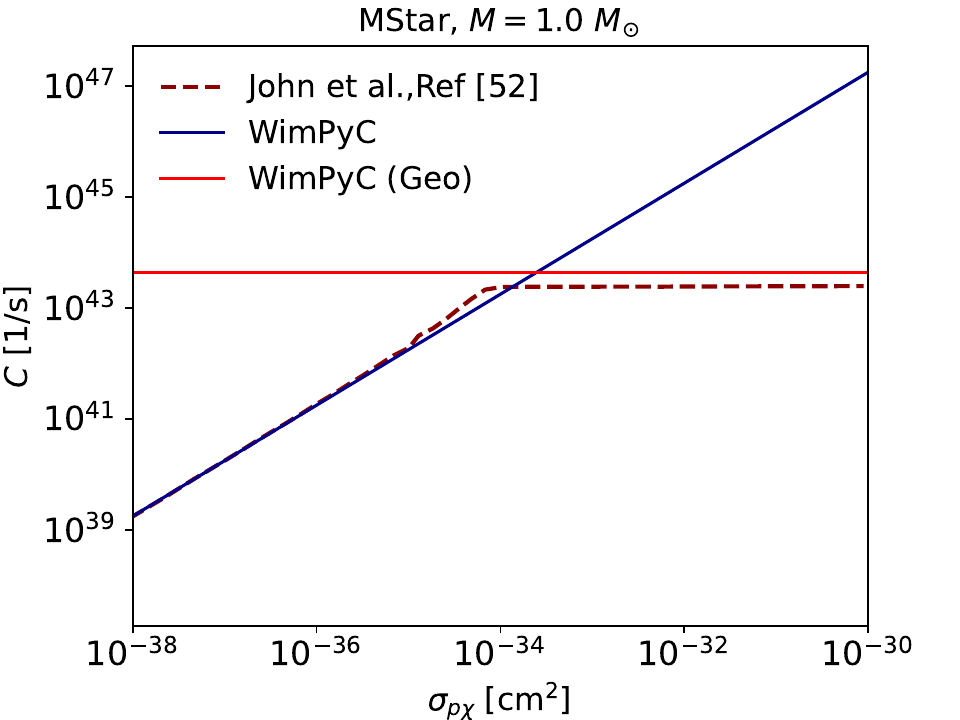}
    \caption{The capture rate calculated using WimPyC for Jupiter (top-left), a light WD (top-right), and a main sequence star (bottom) is compared with the results from Refs.~\cite{asteria}, \cite{leane_wds} and~\cite{leane_ds}, respectively.}
    \label{fig:jupiter_wds_ds}
\end{figure}

In the following example a standard SD interaction (corresponding to the ${\cal O}_4$ operator) is parameterized in terms of the corresponding WIMP--proton cross section $\sigma_{\chi p}$ implemented by the \verb|WD.c_tau_SD| routine (which is the analogous of \verb|WD.c_tau_SI| for a SD interaction). In the top-left panel of Fig.~\ref{fig:jupiter_wds_ds} the \verb|WimPyC| output for Jupiter, $\sigma_{\chi p}=10^{-35}$cm$^2$ and a standard Maxwellian velocity distribution is compared with the result of Refs.~\cite{asteria, Robles:2024tdh}.

\begin{Verbatim}[frame=single,xleftmargin=0.5cm,xrightmargin=0.5cm,commandchars=\\\{\}]
vmin,delta_eta=WD.streamed_halo_function(v_esc_gal=550) \\
mchi_vec=np.logspace(np.log10(1e-1), np.log10(10), 50)
SD=WD.eft_hamiltonian('SD',  \{4: WD.c_tau_SD\})\\
capture=[WD.wimp_capture(WD.Jupiter, SD, vmin, delta_eta, 
mchi, rho_loc=0.4, sigma_p=1e-35, cn_over_cp=0) for mchi 
in mchi_vec]
\end{Verbatim}

\subsection{White Dwarf close to Galactic Center, SI interaction}

In the example below, a White Dwarf close to the Galactic center is considered, following Ref.~\cite{leane_wds}, with parameters: $\rho_\chi\simeq$ 9322.7 GeV cm$^{-3}$, and a Maxwellian distribution where the escape velocity is neglected. The velocity dispersion is $v_{rms}$ = 270 km/s, and the velocity of the White Dwarf in the Galactic rest frame is $\vec{v}_{rot,Gal}$ = 285 km/s. 
\begin{Verbatim}[frame=single,xleftmargin=0.5cm,xrightmargin=0.5cm,commandchars=\\\{\}]
##the escape velocity is neglected, v_esc_gal set to 
large number 
vmin, delta_eta=WD.streamed_halo_function(v_rot_gal=
np.array([ 0., 285., 0.]), v_sun_rot=
np.array([ 0., 0., 0.]), v_esc_gal=10000, vrms=270)
mchi_vec=np.logspace(np.log10(1e-2), np.log10(100), 50)
model=WD.eft_hamiltonian('SI',\{ 1: WD.c_tau_SI\})
capture=[WD.wimp_capture(WD.White_Dwarf, model,vmin, 
delta_eta, mchi, rho_loc=9322.7, sigma_p=1e-35) for mchi 
in mchi_vec]
\end{Verbatim}

In the top-right panel of Fig.~\ref{fig:jupiter_wds_ds} the capture rate calculated using \verb|WimPyC| is compared with the published results from Ref.~\cite{leane_wds}.

\subsection{Main Sequence star, SI interaction, geometric limit}

In the following example \verb|WD.MS_Star| represents a main sequence star with mass and radius of the Sun and 100\% Hydrogen composition close to the Galactic Center. 
\begin{Verbatim}[frame=single,xleftmargin=0.5cm,xrightmargin=0.5cm,commandchars=\\\{\}]
vmin,delta_eta=WD.streamed_halo_function(v_rot_gal=
np.array([0.,1000.,0.]), v_esc_gal=550)
sigma_p_vec=np.logspace(np.log10(1e-39), np.log10(1e-30),
50)
SI=WD.eft_hamiltonian('SI',\{ 1: WD.c_tau_SI\})
capture=[WD.wimp_capture(WD.MS_Star, SI,vmin, delta_eta, 
mchi=0.1, rho_loc=2e12, sigma_p=sigma_p) \\
for sigma_p in sigma_p_vec]
c_geom=WD.wimp_capture_geom(WD.MS_Star, mchi=0.1, vmin=
vmin, delta_eta=delta_eta, rho_loc=2e12)
\end{Verbatim}

In the bottom plot of Fig.~\ref{fig:jupiter_wds_ds} the \verb|WimPyDD| result for the capture rate is plotted as a function of the WIMP--nucleon scattering cross section and compared to that of Ref.~\cite{leane_ds} for a SI interaction, $m_\chi$ = 0.1 GeV, $\rho_\chi$ =2$\times 10^{12}$ GeV cm$^{-3}$ and an orbital velocity of 1000 km/s.  In the same plot the horizontal solid line represents the geometric capture rate calculated by \verb|wimp_capture_geom|. In such plot one can observe the transition from the optically thin to the optically thick regime, and how in \verb|WimPyC| the more refined treatment of Ref.~\cite{leane_ds} can be reproduced reasonably well by interpolating between the optically thin result and the geometric limit.

\subsection{Matching of non–relativistic Wilson coefficients: anapole DM}
\label{subsec:anapole}
The completeness of the base of the NREFT Hamiltonian of Eq.~(\ref{eq:H}) implies that \verb|WimPyC| can be used to calculate the capture rate for any high--energy physics interaction between the WIMP and the nucleus, once the low--energy Wilson coefficients are properly matched. We exemplify this procedure in the specific example of a DM particle that interacts with ordinary matter entirely through an electromagnetic anapole moment~\cite{anapole_Ho_2012, anapole_Fitzpatrick_2010}. The interaction Lagrangian is given by:
\begin{equation}
{\cal L} =\frac{1}{2}\frac{g}{\Lambda^2}\, \bar{\chi}\gamma^{\mu}\gamma^5\chi \, \partial^{\nu}F_{\mu\nu},
  \label{eq:adm_lagrangian}
\end{equation}
\noindent where $\chi$ is the Anapole DM (ADM) field, $F_{\mu\nu}$ is the electromagnetic field strength tensor,  $g$ is a dimensionless coupling constant, and $\Lambda$ is a new physics scale. 
The nonrelativistic scattering of an 
ADM particle with a nucleon can also be described by the contact interaction Hamiltonian
\begin{equation}
  H_{\chi N} = \frac{2e g}{\Lambda^2} \,  \vec{S}_{\chi}\cdot  \bigg (  e_N \vec{v}_{\chi N}^{\perp} +i \frac{(g_N/2)}{m_N} \, \vec{q}\times
    \vec{S}_N \bigg ).
  \label{eq:nr_interaction}
\end{equation}

\noindent In the equation above $e$ is the elementary electric charge, $e_N$ is the nucleon charge in units of $e$ ($e_{\rm p}=1$, $e_{\rm n}=0$), $g_N/2$ is the nucleon
magnetic moment in units of nuclear magnetons $e/2m_N$ ($g_{\rm p}=5.585\,694\,713(46)$, $g_{\rm n}=-3.826\,085\,45(90)$), $m_N$ is the nucleon mass, $\vec{S}_{\chi}$ and
$\vec{S}_{N}$ are the spins of the WIMP and the nucleon, respectively, $\vec{q}$ is the
 momentum transfer, and $\vec{v}^{\perp}_{\chi N}$ is the component of the WIMP--nucleon relative velocity perpendicular to $\vec{q}$. In the base of Ref.~\cite{haxton1} Eq.~(\ref{eq:nr_interaction}) reads:

\begin{equation}
H_{\chi N} = \sum_{\tau=0,1} ( c_8^\tau {\cal O}_8+ c_9^\tau {\cal O}_9) t^\tau,
\end{equation}

\noindent with the $c_8^{\tau}$ and $c_9^{\tau}$ Wilson coefficients given by:

\begin{equation}
c_8^\tau = \frac{2eg}{\Lambda^2} \, e^\tau,
\qquad
c_9^\tau = - \frac{eg}{\Lambda^2} \, g^\tau,
\end{equation}
\noindent and $e^0=e^1=1$, $g^0=g_{\rm p}+g_{\rm n}$, $g^1=g_{\rm p}-g_{\rm n}$. 
In particular, the low--energy phenomenology depends on the WIMP mass $m_\chi$ and on the reference cross section:

\begin{equation}
\label{eq:ADMsigref}
\sigma_{\rm ref} \equiv \frac{2 \mu_{\chi {\cal N}}^2 \, \alpha g^2 }{ \Lambda^4}\, ,
\end{equation}

\noindent where $\alpha = e^2 / 4 \pi \simeq 1/137$ is the fine structure constant and $\mu_{\chi {\cal N}}$ the WIMP--nucleon reduced mass.
The implementation of the model in \verb|WimPyC| is straightforward, and requires to define two functions \verb|c8_anapole(mchi,sigm|
\verb|a_ref)| and \verb|c9_anapole(mchi,sigma_ref)| returning as a function of $m_\chi$ and $\sigma_{\rm ref}$ the 2--dimensional arrays $\left [ c_8^0, c_8^1 \right ]$ and $\left [ c_9^0, c_9^1 \right ]$  containing 
the isoscalar and isovector components of the Wilson coefficients in units of GeV$^{-2}$.
Explicitly:

\begin{Verbatim}[frame=single,xleftmargin=0.5cm,xrightmargin=0.5cm,commandchars=\\\{\}]
def c8_anapole(mchi,sigma_ref):
      hbarc=1.972e-14 #hbar*c in GeV*cm
      e_tau=np.array([1,1])
      m_N=0.931 #nucleon mass in GeV
      m_chi_N=mchi*m_N/(mchi+m_N)
      eg_over_lambda2=np.sqrt(2*np.pi*sigma_ref)/(m_chi_N
      *hbarc)
      return 2*eg_over_lambda2*e_tau

def c9_anapole(mchi,sigma_ref):
      hbarc=1.972e-14 #hbar*c in GeV*cm
      gp=5.585694713
      gn=-3.82608545
      g_tau=np.array([gp+gn,gp-gn])
      m_N=0.931 #nucleon mass in GeV
      m_chi_N=mchi*m_N/(mchi+m_N)
      eg_over_lambda2=np.sqrt(2*np.pi*sigma_ref)/(m_chi_N
      *hbarc)
      return -eg_over_lambda2*g_tau
      
wilson_coeff_anapole=\{8: c8_anapole, 9: c9_anapole\}
anapole=WD.eft_hamiltonian('anapole',wilson_coeff_anapole)
\end{Verbatim}

\begin{figure}[h!]
    \centering
    \includegraphics[width=0.6\columnwidth]{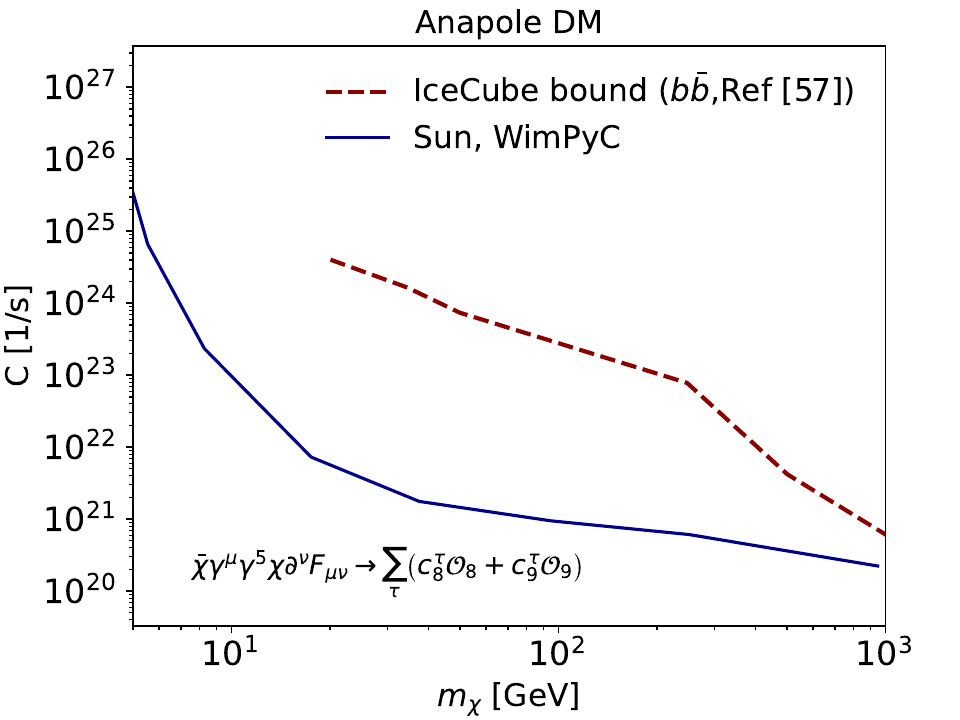}
    \caption{Maximal capture rate in the Sun for Anapole DM allowed by the DD exclusion plot of Ref.~\cite{sogang_gondolo_anapole} compared to the IceCube upper bound~\cite{IceCube_ICRC2021} assuming $b\bar{b}$ annihilation}.
    \label{fig:anapole}
\end{figure}
The largest ADM capture $C_\odot$ in the Sun allowed by DD can be obtained by calculating it over the exclusion plot of $\sigma_{\rm ref}$ vs. $m_\chi$ provided in Fig. 2 of Ref.~\cite{sogang_gondolo_anapole}. In Fig.~\ref{fig:anapole} the corresponding output of \verb|wimp_capture| is compared to the upper bound on the annihilation rate $\Gamma_\odot$ for the $b\bar{b}$ channel~\cite{IceCube_ICRC2021} (see for instance Fig. 5 of~\cite{halo_independent_sogang_long_range}).

\begin{Verbatim}[frame=single,xleftmargin=0.5cm,xrightmargin=0.5cm,commandchars=\\\{\}]
# mchi and sigma_ref from exclusion plot of 1808.04112
mchi_lim=np.array([4.35,4.60,5.54,8.26,17.52,37.49,94.80,
251.07,945.86])
sigma_ref_lim=np.array([2.36e-34,1.91e-35,1.10e-36, 
4.92e-38,2.54e-39,1.20e-39,1.81e-39,4.29e-39,1.42e-38])
#standard Maxwellian
vmin, delta_eta=WD.streamed_halo_function() 
#calculation of ADM capture rate in the Sun
capture_vec=np.array([])
for mchi,sigma_ref in zip(mchi_lim,sigma_ref_lim):
    capture=WD.wimp_capture(WD.Sun, anapole, 
    vmin,delta_eta, mchi,sigma_ref=sigma_ref) 
    capture_vec=np.append(capture_vec,capture)
#check equilibrium between capture and annihilation
annihilation_vec=np.array([])
for mchi,capture in zip(mchi_lim,capture_vec):
    annihilation=WD.wimp_capture_annihilation(WD.Sun,mchi,
    capture)
    annihilation_vec=np.append(annihilation_vec,
    annihilation)
print(capture_vec/(2*annihilation_vec))
[1. 1. 1. 1. 1. 1. 1. 1. 1.]
\end{Verbatim}
As shown at the bottom of the example, using the \verb|wimp_capture_annihilati|
\verb|on| routine (see \ref{app:annihilation}) one can numerically verify that the equilibrium condition $\Gamma_\odot=C_\odot/2$ holds in the full range of $m_\chi$ (when the default value $\langle \sigma v\rangle$ = 3$\times$ 10$^{-26}$ cm$^3$ s$^{-1}$ is used for the WIMP annihilation cross section times velocity). 

\begin{figure}[h!]
    \centering
    \includegraphics[width=0.45\textwidth]{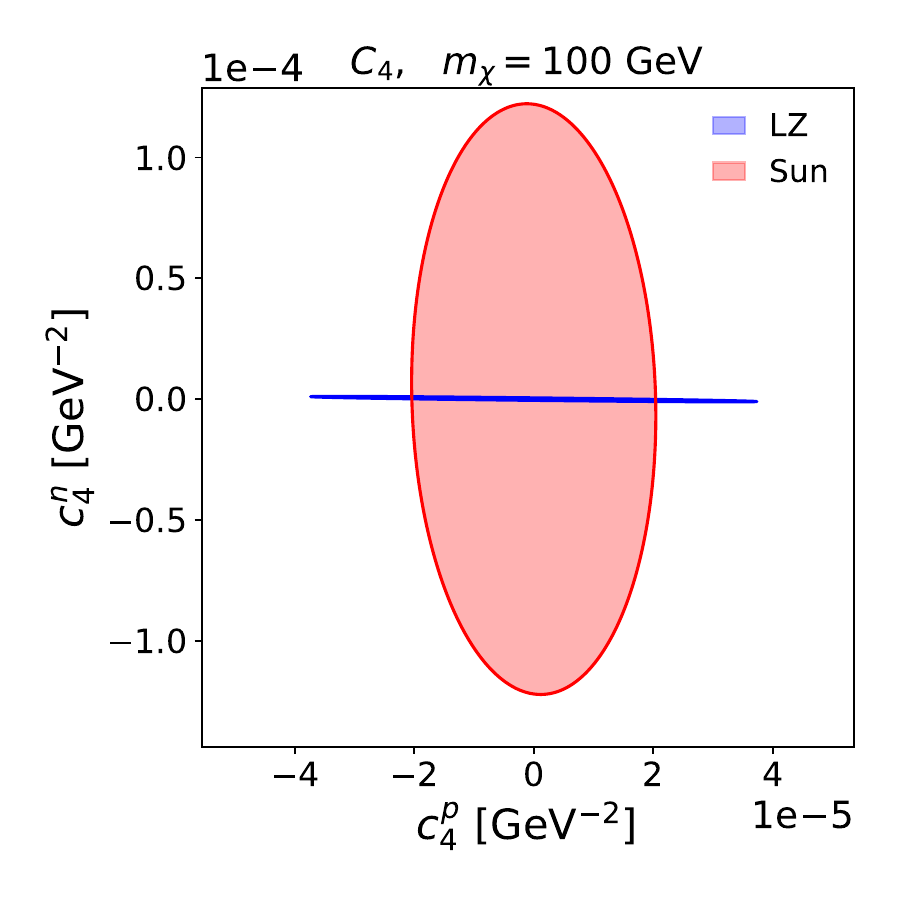} \hspace{-0.3cm}
    \includegraphics[width=0.46\textwidth]{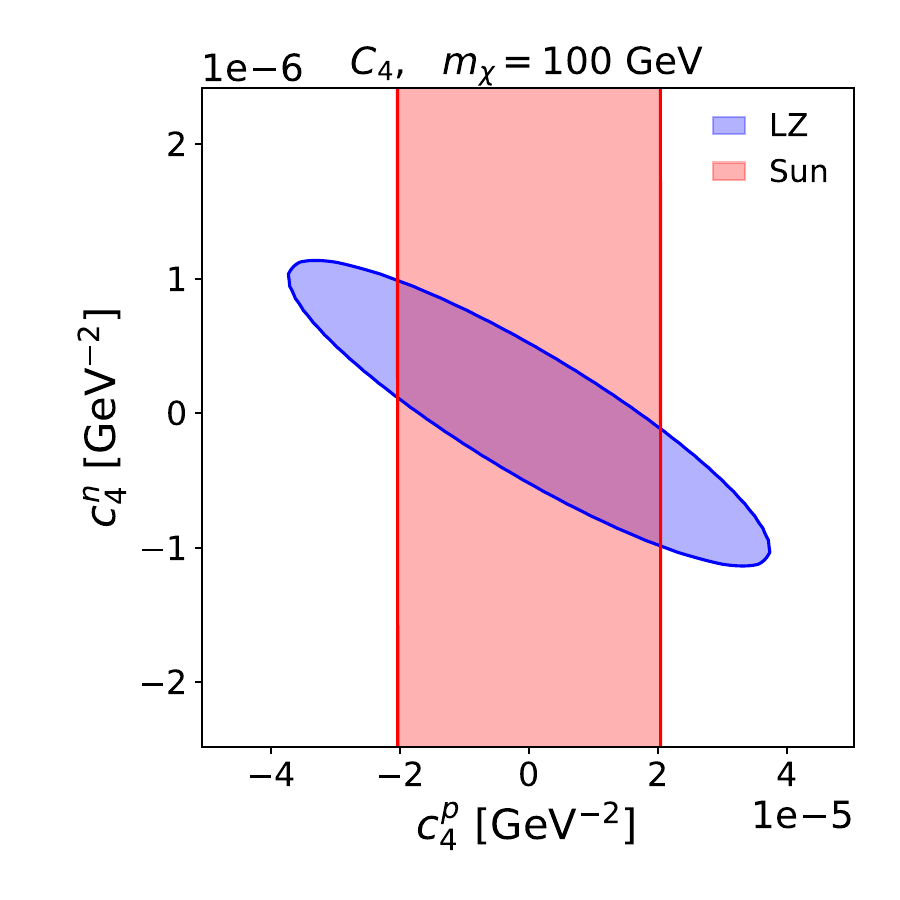} 
    \caption{$90\%$ C.L. allowed region in the $c^p \times c^n$ plane for the DD experiment LZ~\cite{LZ_DD_bound_2025}, and considering capture inside the Sun while assuming $b\bar{b}$ annihilation for the IceCube upper bound~\cite{IceCube_ICRC2021}. The left panel corresponds to the $\mathcal{O}_4$ (standard SD) interaction, while the  right panel is its zoomed-in version.}
    \label{fig:ellipses}
\end{figure}
\subsection{Matrix in couplings space for the calculation of the capture rate}
\label{subsec:matrix}
The effective couplings of the Anapole model of the previous example lie in the four--dimensional vector space $c=(c_8^0,c_8^1, c_9^0, c_9^1)$. For a specific value of the WIMP mass \verb|mchi| the corresponding matrix ${\cal M}$ can be obtained using \verb|wimp_capture_matrix|. In such case {\it an auxiliary effective Hamiltonian with all couplings fixed to unity needs to be defined} (values different from unity can be used, if a different normalization of the couplings and of the matrix elements is needed). Notice that the class \verb|eft_hamiltonian| always expects that the Wilson coefficient is a function, so in this case the routine with no arguments \verb|lambda: [1,1]| is used to parameterize constant coefficients.

\begin{Verbatim}[frame=single,xleftmargin=0.5cm,xrightmargin=0.5cm,commandchars=\\\{\}]
# auxiliary model needed by wimp_capture_matrix
wilson_coeff_aux=
\{k:lambda:[1,1] for k in wilson_coeff_anapole.keys()\}
anapole_aux=WD.eft_hamiltonian('anapole_aux',
wilson_coeff_aux)
# use mchi_lim[6]=94.8 from previous example
matrix=WD.wimp_capture_matrix(WD.Sun,anapole_aux,vmin,
delta_eta,mchi_lim[6])
matrix.shape
(4, 4)
#mapping between couplings and positions in the matrix
WD.get_mapping(anapole_aux)
{(8, 0): 0, (8, 1): 1, (9, 0): 2, (9, 1): 3}
#couplings vector using sigma_ref_lim[6]=1.81e-39 
c_vec=np.append(c8_anapole(mchi_lim[6],sigma_ref_lim[6]),
c9_anapole(mchi_lim[6],sigma_ref_lim[6]))
[1.17426826e-05, 1.17426826e-05, -1.03312665e-05,
-5.52597735e-05]
# capture rate 
capture_matrix=np.dot(c_vec,np.dot(matrix,c_vec))
9.490518349177376e+20
\end{Verbatim}

The capture rate calculated in this way matches the output of \verb|wimp_captu|
\verb|re| in the previous example.

In particular, as shown in Ref.~\cite{bracketing_sogang_ibarra_2022}, the complementarity of DD experiments and WIMP capture in the Sun can be studied using the matrix-based formalism, where the parameter space allowed by each experiment $exp$ lies inside the ellipsoid $c^T \cdot {\cal M}_{exp}\cdot c< R_{exp}^{max}$, with $R_{exp}^{max}$ the corresponding experimental upper bound. In Fig.~\ref{fig:ellipses}, we compare the latest results of the LZ experiment~\cite{LZ_DD_bound_2025} with the IceCube upper bound assuming $b\bar{b}$ annihilation~\cite{IceCube_ICRC2021}, by plotting the allowed parameter space for each experiment in the $c^p \times c^n$ plane for $m_{\chi}=100$ GeV, assuming the $\mathcal{O}_4$ interaction. The explicit implementation of the relevant matrices, which also makes use of the \verb|wimp_dd_matrix| routine for DD, corresponds to:
\begin{Verbatim}[frame=single,xleftmargin=0.5cm,xrightmargin=0.5cm,commandchars=\\\{\}]
vmin,delta_eta=WD.streamed_halo_function()
c4=WD.eft_hamiltonian('c4',\{4: lambda: [1,1]\})
U_pn= WD.rotation_from_isospin_to_pn(c4)
matrix_Sun = WD.wimp_capture_matrix(WD.Sun, c4, vmin, 
delta_eta, mchi=100)
matrix_Sun_pn = np.dot(U_pn.T, np.dot(matrix_Sun, U_pn))
LZ=WD.experiment('LZ_2022')
matrix_LZ=WD.wimp_dd_matrix(LZ,c4,0,vmin,delta_eta, mchi=
100)
matrix_LZ_pn = np.dot(U_pn.T, np.dot(matrix_LZ, U_pn))
\end{Verbatim}
\noindent Notice that the output of both \verb|wimp_capture_matrix| and \verb|wimp_dd_matrix| is in isospin space and needs to be rotated into the $(p,n)$ space using the output of\\ \verb|rotation_from_isospin_to_pn|.

\subsection{Halo--independent methods}
\label{sec:halo_independent}

The routine \verb|wimp_capture| takes as inputs the arrays \verb|vmin|, \verb|delta_eta|  containing the halo function and 
returns the capture rate by summing the  master formula of Eq.~(\ref{eq:capture_master}) over the velocity streams 
contained in \verb|vmin|, \verb|delta_eta|. To control this behavior \verb|wimp_capture| has the argument \verb|sum_ov|
\verb|er_streams| with default value \verb|True|. If, instead,  \verb|sum_over_streams=False| the sum over the velocity streams is not performed and \verb|wimp_capture| returns an array with 
shape \verb|vmin.shape| (in particular \verb|wimp_capture| accepts any shape for \verb|vmin| and \verb|delta_eta|,
allowing to sample a large number of stream sets efficiently with a single call).  

\begin{figure}[h!]
    \centering
    \includegraphics[width=0.6\textwidth]{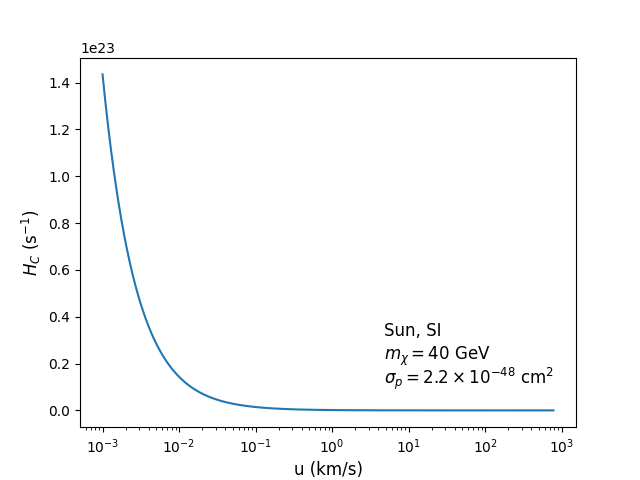} 
    \caption{Response function  ${\cal H}$ as a function of $u$ for a SI interaction and $m_\chi$ =40 GeV, $\sigma_p$ = 2.2$\times$ 10$^{-48}$ cm$^2$ (corresponding to the LZ upper bound from DD~\cite{LZ})}.
    \label{fig:Hc_u}
\end{figure}

This feature allows to calculate the response function ${\cal H}_C(u)$ defined in Eq.~(\ref{eq:lambda_k}). In particular ${\cal H}_C(u_k)$ can be factored out from the capture rate (see Eq.~(\ref{eq:Hk})) by setting $\lambda_k=\delta\eta_k u_k=1$, i.e. $\delta\eta_k=1/u_k$:

\begin{Verbatim}[frame=single,xleftmargin=0.5cm,xrightmargin=0.5cm,commandchars=\\\{\}]
u=np.logspace(-3, np.log10(782), 1000)
h_c=WD.wimp_capture(WD.Sun, WD.SI, u, 1/u, mchi=40,
sum_over_streams=False,sigma_p=2.2e-48)
pl.plot(u,h_c)
\end{Verbatim}

In the example above the WIMP LZ bound~\cite{LZ_DD_bound_2025} on the WIMP--proton cross $\sigma_p$ for a SI interaction at $m_\chi$ = 40 GeV (2.2$\times$ 10$^{-48}$ cm$^2$) is used to calculate the response function ${\cal H}$ as a function of $u$. The output is shown in Fig.~\ref{fig:Hc_u}.

The way \verb|wimp_capture| can be used for halo--independent methods to calculate the response function ${\cal H}_C(u)$ is identical to that of the \verb|wimp_dd_rate| routine for DD, described in Sections 2.1 and 3.5.4 of~\cite{WimPyDD_2022}.

\subsection{Massless propagator}
\label{subsec:massless_propagator}
Since the Wilson coefficients of the effective Hamiltonian Eq.~(\ref{eq:H}) are arbitrary functions of the transferred momentum $q$ in the functions passed to the \verb|eft_hamiltonian| class the argument \verb|q| is reserved for the transferred momentum $q$ in GeV\footnote{The other two reserved arguments are \verb|mchi| and \verb|delta| for the WIMP mass $m_\chi$ in GeV and the inelastic mass splitting $\delta$ in keV, respectively (see~\cite{WimPyDD_2022} for a detailed description).}.  A common situation is when a momentum dependence is induced by a massless propagator.  In this case the capture rate formally diverges, because it is driven by WIMPs locked on very large bound orbits with $q\rightarrow$ 0, where the WIMP--nucleus cross section diverges. A cut-off on the capture rate is usually obtained by assuming that the gravitational disturbances far from the Sun put an upper cut $r_0$ on the size of the maximal aphelion that a captured WIMP can have~\cite{Peter_Jupiter_cut_2009}. In particular, indicating with $R_\odot$ and $v_{esc}(R_\odot)$ the Sun's radius and the escape velocity from the Sun's surface, a WIMP with initial position $r$ needs a minimal speed $v_e(r)^2=v_{esc}(r)^2-v_{esc}(r_0)^2$ with $v_{esc}(r_0)^2=v_{esc}(R_\odot)^2R_\odot/r_0$ to reach the maximal distance $r_0$. In this case the expression of Eq.~(\ref{eq:E_cap}) for the minimal energy required for capture is modified to: 

\begin{equation}
    E^\chi_{\rm cap}(u)  \rightarrow \frac{1}{2} m_\chi (u^2 + v_{esc}(r_0)^2) -\delta
    \label{eq:jupiter_cut}
\end{equation}

\noindent in order for the WIMP to be locked into a bound orbit with aphelion $r_0$.  In the literature, $r_0$ is identified with the distance from the Sun to Jupiter~\cite{jupiter_cut_Kumar_2012,Liang:2013dsa, Guo:2013ypa}, corresponding to:

\begin{equation}
v_{\rm esc}(r_0) = v_{\rm cut} \simeq 18.5 \, {\rm km/s} \, .
\label{eq:v_cut}
\end{equation}
\noindent A non-vanishing value of $v_{\rm cut}$ increases $E^\chi_{\rm cap}$, making the condition for capture more stringent.  It can be passed to the \verb|wimp_capture| routine through the argument \verb|v_cut|. Following Ref.~\cite{Guo:2013ypa}, in the example below the Jupiter cut is applied for an isoscalar SI WIMP-nucleon interaction with $\sigma_{\rm ref}$ = $\sigma_p$ = $\sigma_n$= $(c_1)^2 \mu^2_{\chi{\cal N}}/\pi\times F^2_{DM}(q)$ with $F_{DM}$ = $q_0/q$ or $(q_0/q)^2$ and $q_0$ = 0.1 GeV. In particular, the desired momentum--dependent Wilson coefficients can be obtained by multiplying the output of \verb|WD.c_tau_SI| by $F^2_{DM}$. Also, the coupling \verb|1| is modified to \verb|(1,'qm1')| for $F_{DM}$ = $q_0/q$ or \verb|(1,'qm2')| for $F_{DM}$ = $(q_0/q)^2$ in order to fix the name of the files where the response functions are saved. For instance, for Iron instead of \verb|56Fe_c_1_c_1.npy|, which contains the response functions for a contact interaction, the file name \verb|56Fe_c_1_qm1_c_1_qm1.npy| is used for $F_{DM}$ = $q_0/q$ and the file name \verb|56Fe_c_1_qm2_c_1_qm2.npy| is used for $F_{DM}$ = $(q_0/q)^2$ (a detailed discussion on how to handle the output files of different sets of response functions is provided in Section 3.4 of Ref.~\cite{WimPyDD_2022}).

\begin{Verbatim}[frame=single,xleftmargin=0.5cm,xrightmargin=0.5cm,commandchars=\\\{\}]
# cn_over_cp = 1 for isoscalar interaction
c1_tau_SI_qm1=lambda sigma_p,mchi,q,cn_over_cp=1,q0=0.1: 
WD.c_tau_SI(sigma_p,mchi, cn_over_cp)*q0/q
c1_tau_SI_qm2=lambda sigma_p,mchi,q,cn_over_cp=1,q0=0.1: 
WD.c_tau_SI(sigma_p,mchi, cn_over_cp)*(q0/q)**2
model_qm1=WD.eft_hamiltonian('F_DM_qm1',\{(1,'qm1'): 
c1_tau_SI_qm1\})
model_qm2=WD.eft_hamiltonian('F_DM_qm2',\{(1,'qm2'): 
c1_tau_SI_qm2\})
\end{Verbatim}

\begin{figure}[h!]
    \centering
    \includegraphics[width=0.6\columnwidth]{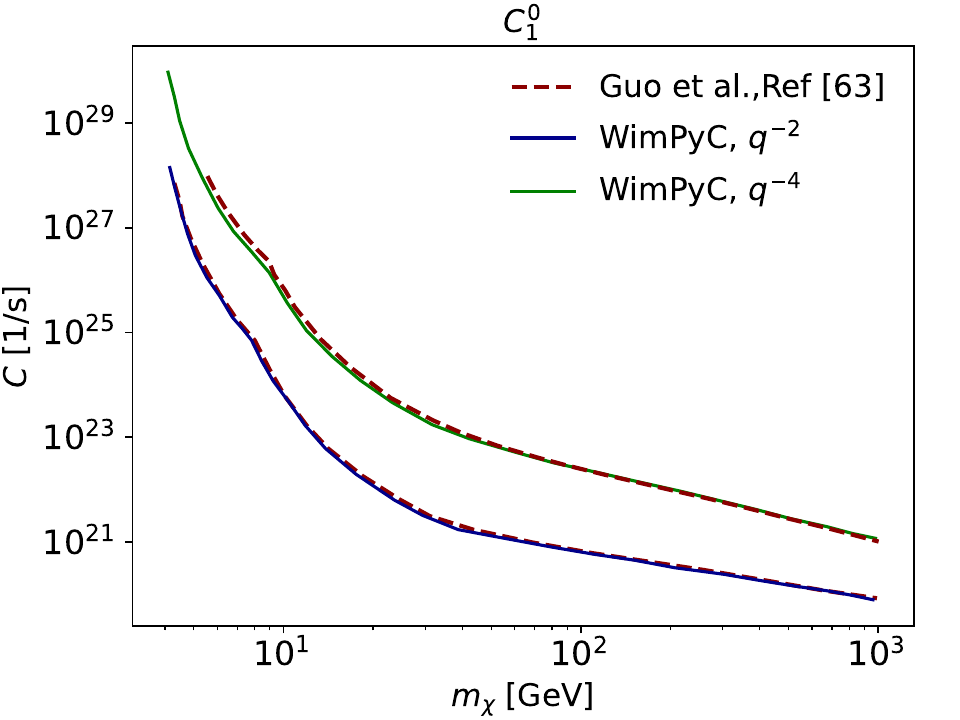}
    \cprotect\caption{The output of \verb|WimPyC| for an isoscalar SI interaction modified by the momentum-dependent term $F_{DM}$ = $q_0/q$ or $(q_0/q)^2$ with $q_0$ = 0.1 GeV is compared to that of Ref.~\cite{Guo:2013ypa} when $m_\chi$ and $\sigma_{\rm ref}$ are taken from the corresponding exclusion plots in Fig. 1 of the same paper. 
    \label{fig:long_range}}
\end{figure}

\noindent Indicating with \verb|mchi| and \verb|sigma_ref| the WIMP mass $m_\chi$ and the reference cross section $\sigma_{\rm ref}$ the capture rate in the Sun, including the Jupiter cut, is then obtained using:    
\begin{Verbatim}[frame=single,xleftmargin=0.5cm,xrightmargin=0.5cm,commandchars=\\\{\}]
capture_qm1=WD.wimp_capture(WD.Sun, model_qm1, vmin, 
delta_eta, mchi=mchi, sigma_p=sigma_ref, v_cut=18.5)
capture_qm2=WD.wimp_capture(WD.Sun, model_qm2, vmin, 
delta_eta, mchi=mchi, sigma_p=sigma_ref, v_cut=18.5)
\end{Verbatim}

In Fig.~\ref{fig:long_range} the output of \verb|WimPyC| is compared to that of Ref.~\cite{Guo:2013ypa} when $m_\chi$ and $\sigma_{\rm ref}$ are taken from the corresponding exclusion plots in Fig. 1 of the same paper.

\subsection{Differential capture}
\label{subsec:diff_capture}
The capture rate of Eq.~(\ref{eq:capture_master}) requires to perform the following radial integral, summed over the velocity streams $u_k$, the nuclear targets $T$ and each $c_ic_j$  combination of couplings in the effective Hamiltonian (Eq.~(\ref{eq:H})):

\begin{equation}
    C_{opt}=\sum_T \sum_{ij}\sum_k\int_0^{R_\odot} d^3 r   \frac{dC^T_{ij}}{d^3 r du}(r,u_k).
    \label{eq:dc_dvi}
\end{equation}

\noindent In particular, $\frac{dC^T_{ij}}{d^3 r du}$ is related to the ${\cal H}_C$ function of Eq.~(\ref{eq:HC}):

\begin{equation}
    {\cal H}_C(u)=\frac{\rho_\chi}{m_\chi}\int_0^{R_\odot} d^3r ~~ 4\pi r^2 \sum_T\sum_{ij}\frac{1}{u}\frac{dC^T_{ij}}{d^3 r du}(r,u).
\end{equation}

\noindent An array containing the values of $\frac{dC^T_{ij}}{d^3 r du}(r,u_k)$ (in s$^{-1}$ cm$^{-3}$), tabulated over the \verb|r_vec| values of the \verb|celestial_body| object and the \verb|vmin| array is accessible through the dynamic attribute \verb|dc_dvi| that the routines \verb|wimp_capture| and  \verb|wimp_capture_accurate| acquire after each call. \verb|dc_dvi| is a python dictionary that accepts as key the \verb|symbol| attribute of the target and one of the coupling combinations in the \verb|coeff_squared_list| attribute of the interaction Hamiltonian. In the example below the WIMP LZ bounds~\cite{LZ_DD_bound_2025} on the WIMP--proton cross $\sigma_p$ for the SI and SD interactions at $m_\chi$ =40 GeV (2.2$\times$ 10$^{-48}$ cm$^2$ and  9.8$\times$ 10$^{-42}$ cm$^2$, respectively) are used to calculate the corresponding capture rates in the Sun, and the dynamic attribute \verb|dc_dvi| is used to access the $\frac{dC^T_{ij}}{d^3 r du}(r,u_k)$ values for the $^{56}Fe$ and the $^{27}Al$ targets.
Notice that in this case the first value of the  \verb|vmin| array is equal to zero and is dropped from the sampling.

\begin{Verbatim}[frame=single,xleftmargin=0.5cm,xrightmargin=0.5cm,commandchars=\\\{\}]
len(WD.Sun.r_vec)
135
vmin,delta_eta=WD.streamed_halo_function()
len(vmin)
1000
capture_SI=WD.wimp_capture(WD.Sun, WD.SI, vmin, delta_eta, 
mchi=40, sigma_p=2.2e-48)
dc_dvi_SI=WD.wimp_capture.dc_dvi['56Fe',(1,1)]
dc_dvi_SI.shape
(135, 999)
capture_SD=WD.wimp_capture(WD.Sun, WD.SD, vmin, delta_eta, 
mchi=40, sigma_p=9.8e-42)
dc_dvi_SD=WD.wimp_capture.dc_dvi['27Al',(4,4)]
dc_dvi_SD.shape
(135, 999)
\end{Verbatim}

\section*{Acknowledgments} The work of S.S.\ and S.K.\ is supported by Basic Science Research Program through the National Research Foundation of Korea (NRF) funded by the Ministry of Education through the Center for Quantum Spacetime (CQUeST) of Sogang University with grant number RS-2020-NR049598 and by the Ministry of Science and ICT with grant number RS-2025-24523022. G.T. acknowledges his visit to Sogang University during which part of this work was done.

\appendix

\section{Initializing the celestial body directory}
\label{app:initializing_celestial_body}

\begin{table}[t]\centering
  \caption{Files in the celestial body subfolder containing the information
    to calculate the capture rate. \label{table:files}}
\renewcommand{\arraystretch}{1.1}
\addtolength{\tabcolsep}{2.0pt}
\vskip\baselineskip
\begin{tabular}{@{}ccc@{}}
\toprule
{\bf Physical quantity(unit)} & {\bf Default(if missing file)}& {\bf File name}  \\
\midrule
Mass (solar masses) & 1 & {\verb|mass.tab|}\\
\midrule
radius (solar radii) & 1 & {\verb|radius.tab|} \\
\midrule
Core temperature (K) & 1.4e7 & {\verb|core_temperature.tab|}\\
\midrule
density profiles & hydrogen sphere  & {\verb|densities.tab|}\\
(arbitrary units)  & of constant density &\\
\bottomrule
\end{tabular}
\label{tab:exp_file_list}
\end{table}

Upon instantiation the \verb|celestial_body| class
reads the relevant information from a sub--folder of \verb|WimPyDD/WimPyC/Celestial_bodies|. The name of the subfolder is passed as an
argument to the \verb|celestial_body| class. For instance, the instruction:

\begin{Verbatim}[frame=single,xleftmargin=0.5cm,xrightmargin=0.5cm,commandchars=\\\{\}]
sun=WD.celestial_body('Sun')
\end{Verbatim}

\noindent reads the files listed in Table~\ref{table:files} searching for them in the folder:

\begin{Verbatim}[frame=single,xleftmargin=0.5cm,xrightmargin=0.5cm,commandchars=\\\{\}]
WimPyDD/WimPyC/Celestial_bodies/Sun
\end{Verbatim}

\noindent The \verb|densities.tab| file contains a table with the density profiles of the target elements in the celestial body.

\begin{Verbatim}[frame=single,xleftmargin=0.5cm,xrightmargin=0.5cm,commandchars=\\\{\}]
# Solar density profiles from model AGSS09ph [arXiv.0909.
2668]  
r_vec    rho_tot     1H        4He     v_esc
0.0015    150.5    74.636    130.656     \\
0.0020    150.5    67.190    117.570      \\
0.0025    150.5    64.525    112.846      \\\
...\\
...\\
...
\end{Verbatim}

In particular, as in the example above the file can start with a first optional line beginning with \verb|#| containing documentation text that is added to the \verb|celestial_body_obj.__doc__| attribute of the \verb|celestial_body_obj| object and printed out by the \verb|print(celestial_body_obj)| command. The table content must be preceded by a line containing headers (without apostrophes) that allow \verb|WimPyC| to identify the columns that follow. In particular the code looks for at least three headers: \verb|r_vec|, corresponding to the radial distance from the center; \verb|rho_tot|, corresponding to the total density profile; the valid symbol of at least one nuclear target (i.e. whose label must match one of the strings that identify a valid isotope object, and whose information is stored in the \verb|WimPyDD/Targets| folder). In the numerical tables the units for the radial distance and for all the densities can be arbitrarily chosen, because, before loading them in the corresponding attributes of the \verb|celestial_body_obj| object, \verb|WimPyC| rescales them using the information contained in the \verb|mass.tab| and \verb|radius.tab| files (which contain a single numerical value, besides an optional documentation line starting by \verb|#|) so that $\int_0^1 4\pi r^2 \rho_{tot}(r)$=1 and $\int_0^1 4\pi r^2 \rho_i(r)$ = $f_i$. Notice that, in order to rescale the radius correctly, \verb|WimPyC| assumes that the last tabulated radius in \verb|densities.tab| corresponds to the radius of the celestial body (i.e. the last tabulated radius in \verb|densities.tab| and the value of the radius contained in \verb|radius.tab| must be consistent). The only quantity in \verb|densities.tab| that is required to be expressed in a specific unit is the escape velocity, corresponding to the \verb|v_esc| header, that, if included, must be tabulated in km/s. If a column with header \verb|v_esc| is initially not included \verb|WimPyC| calculates it in km/s using the available information. In such case the user is prompted to add the new column with the tabulated escape velocity to the \verb|densities.tab| file (this avoids to recalculate the escape velocity at every new instantiation).

For testing purposes and in order to provide a template folder that is easy to adapt to specific needs, \verb|WimPyC| allows to instantiate a celestial body without initializing the corresponding folder in the  \verb|WimPyDD/WimPyC/Celestial_b|
\verb|odies| directory, or when the folder is empty. In particular, if the \verb|WimPyDD/Wim|
\verb|PyC/Celestial_bodies/Template| folder does not exist, the instruction:

\begin{Verbatim}[frame=single,xleftmargin=0.5cm,xrightmargin=0.5cm,commandchars=\\\{\}]
template=WD.celestial_body('Template',verbose=True)\\
\end{Verbatim}
\noindent creates it, with the four files \verb|core_temperature.tab|, \verb|mass.tab|, \verb|radius.tab| and \verb|densities.tab| that initialize by default a hydrogen sphere of unit mass and unit radius in solar units (at command line the user can modify both the target, the size and the mass):
\begin{Verbatim}[frame=single,xleftmargin=0.5cm,xrightmargin=0.5cm,commandchars=\\\{\}]
Missing folder WimPyDD/WimPyC/Celestial_bodies/Template/ \\
created \\
densities.tab is missing in \\
WimPyDD/WimPyC/Celestial_bodies/Template/\\
Sphere of constant density with single nuclear target 
will be created\\
choose target (default: 1H)\\
Please input the celestial body mass in solar units. \\
Default value: 1 \\
saved mass 1 in \\
WimPyDD/WimPyC/Celestial_bodies/Template/mass.tab\\
Please input the celestial body radius in solar units. \\
Default value: 1\\
saved radius 1 in \\
WimPyDD/WimPyC/Celestial_bodies/Template/radius.tab\\
Please input the celestial body core_temperature 
in Kelvin. Default value: 1.4e7\\
saved core_temperature 1.4e7 in \\
WimPyDD/WimPyC/Celestial_bodies/Template/
core_temperature.tab\\
WARNING: v_esc label not found in densities.tab file. \\
Check spelling if appropriate. Calculating v_esc from \\
total density\\
Want to add the escape velocity to \\
WimPyDD/WimPyC/Celestial_bodies/Template/densities.tab? \\
(yes,no, default=no)\\
yes\\
looking for 1H in list of pre-defined targets:\\
1H found in predefined hydrogen element object \\
\end{Verbatim}

\section{Celestial Bodies Implemented in the WimPyC distribution}
\label{app:predefined_celestial_bodies}

We briefly summarize here the ingredients used to initialize the  celestial bodies pre-defined in \verb|WimPyC|:

\begin{itemize}
    \item {\bf Sun:} To initialize \verb|WD.Sun| we implemented the density profiles of the Standard Solar Model AGSS09ph~\cite{Serenelli:2009yc}.   In the case of the Sun the capture rate drops to zero due to the effect of evaporation~\cite{evaporation} for WIMP masses lighter than about 4 GeV. As already pointed out \verb|WimPyC| does not calculate such effect, so its output should not be trusted for WIMP masses approaching such lower bound. 
    \item {\bf Earth:} For the implementation of the density profile of \verb|WD.Earth|, we use the Preliminary Reference Earth Model (PREM) of Ref.~\cite{prem}, with the chemical composition of the Earth taken from Table I of Ref.~\cite{bramante_earth}. In the case of the Earth the evaporation effect is expected to become significant for DM particles lighter than approximately 10 GeV~\cite{evaporation}.
    \item{\bf Jupiter:} In WimPyC the density profile of the \verb|WD.Jupiter| object uses the Jovian model J11-4a from Ref.~\cite{jupiter_profile}. For its composition, we assumed $75\%$ hydrogen and $25\%$ helium.  Depending on the scattering cross section the DM evaporation mass for Jupiter lies in the range of approximately 200 MeV to 1 GeV~\cite{leane_jupiter}.
    \item {\bf White Dwarf:} The \verb|WD.White_Dwarf| object represents a White Dwarf with 0.49 $M_\odot$ mass and $9.39\times 10^3$ km radius. We adopted the density profile from Ref.~\cite{improved_WD_2021} assuming a 100\% carbon composition. 
    For WDs the evaporation effect becomes significant for DM particles $\lesssim 1$ MeV~\cite{evaporation}.
    \item {\bf Main sequence star:} In \verb|WimPyC| the \verb|WD.MS_Star| object represents the very simple implementation of a Main Sequence (MS) star: a $100\%$ hydrogen star with mass and radius of the Sun and the density profile of the Standard Solar Model AGSS09ph~\cite{Serenelli:2009yc}. In Refs.~\cite{John:2023knt, leane_ds} it is pointed out that 
    in stars close to the GC the dark matter density can be high enough for dark matter annihilation to substantially replace nuclear fusion (solving the `paradox of youth'~\cite{youth_paradox_Hassani_2020} according to which such stars are young but show spectroscopic features of the more evolved ones). In such case the evaporation effect can be neglected down to very low WIMP masses, as in the example of Fig.~\ref{fig:jupiter_wds_ds}. 

\end{itemize}

\section{Annihilation rate}
\label{app:annihilation}

The \verb|WimPyC| code does not address issues such as the thermalization of the WIMPs inside the celestial body~\cite{thermalization_Griest_1986} as well as the connected evaporation effect~\cite{evaporation}, which is expected to suppress the capture rate at small WIMP masses (for this reason, for each of the celestial bodies listed in \ref{app:predefined_celestial_bodies} and included in the code distribution we provide some information about the minimal WIMP mass below which the \verb|WimPyC| calculation is not supposed to be valid due to evaporation). 

In order to calculate the WIMP annihilation rate under some standard assumptions the \verb|WimPyC| module contains the routine \verb|wimp_capture_annihilat|
\verb|ion|:

\begin{Verbatim}[frame=single,xleftmargin=0.5cm,xrightmargin=0.5cm,commandchars=\\\{\}]
WD.wimp_capture_annihilation(celestial_body_obj, mchi,
capture_rate, sigma_v=3e-26, effective_volume=None,
t_age=4.603e9)
\end{Verbatim}

The routine takes as an input the capture rate in $s^{-1}$, the WIMP mass in GeV, while the arguments \verb|sigma_v| and \verb|t_age| represent, respectively, the thermally averaged WIMP annihilation cross-section $\langle \sigma v\rangle$ in cm$^3$s$^{-1}$ (set by default to the standard value for a thermal relic \verb|sigma_v=3e-26|) and the age of the celestial body in years (set by default to that of the solar system, $t_{age}$ = 4.6$\times 10^9$ years).

Following the standard treatment~\cite{thermalization_Griest_1986} and neglecting evaporation the WIMP annihilation rate is calculated using:

\begin{equation}
    \Gamma = \frac{C_{opt}}{2} \tanh^2 (\frac{t_{age}}{\tau})
    \label{eq:gamma_rate}
\end{equation}

\noindent where the equilibration time $\tau$ is given by:

\begin{equation}
\tau= (C_{opt} C_A)^{-\frac{1}{2}},     
\label{eq:t_eq}
\end{equation}

\noindent with:

\begin{eqnarray}
    &&C_A=\frac{\langle \sigma v \rangle}{V_{eff}}\label{eq:ca}.
    \label{eq:CA}
\end{eqnarray}
\noindent To calculate the effective volume $V_{eff}$ the routine uses by default the standard treatment of Ref.~\cite{thermalization_Griest_1986}, i.e.:

\begin{equation}
\label{eq:v_eff}
    V_{eff}=\frac{\left [4\pi \int_{0}^{R} r^2 n_\chi(r)\,dr\right ]^2}{4\pi \int_{0}^{R} r^2 n_\chi^2(r)\,dr},
\end{equation}
\noindent with $n_\chi$ the WIMP density profile, given by:

\begin{eqnarray}
\label{eq:n_chi}
    n_\chi(r)&=&\frac{1}{\pi^{\frac{3}{2}}r_\chi^3} e^{-\frac{r^2}{r_\chi^2}},\nonumber \\
    r_\chi^2&=&\frac{3 k T_c}{2\pi m_\chi G\rho_c},
\end{eqnarray}
\noindent assuming full thermalization. In the equation above $k$ is the Boltzmann constant, $G$ the gravitational constant, $\rho_c$ the core mass density and $T_c$ the central temperature of the celestial body (in this case the \verb|T_c| attribute of the \verb|celestial_body| object is used).

When WIMPs do not thermalize with the celestial body core, or in case of a partial thermalization,  the \verb|effective_volume| argument (in cm$^3$) can be used to provide an alternative evaluation of the effective volume of Eq.~(\ref{eq:v_eff}). When the thermalization process is poorly known the conservative case of minimal signal corresponds to a constant WIMP density $n_\chi$,  for which $V_{eff}$ reaches it maximal value equal to the volume of the celestial body. Interestingly, in the case of the Sun this relaxes the present bounds from neutrino detectors by only a factor of a few~\cite{halo_independent_sogang_inelastic}.
When in Eq.~(\ref{eq:t_eq}) the combination $C_{opt}C_A$ is large enough 
one has $\tau\ll t_{age}$ and in Eq.~(\ref{eq:gamma_rate}) the annihilation rate does not depend on $\langle \sigma v\rangle$, saturating to its equilibrium value $\Gamma=C/2$.

\section{Accuracy of the interpolation method}
\label{app:accuracy}

The interpolation method used by the \verb|wimp_capture| routine speeds-up the computational time by more than 2 orders of magnitude compared to  \verb|wimp_capture_accurate|, which makes use of no interpolation. This can make a significant difference, especially when running \verb|WimPyC| on a laptop. As a consequence, as already explained, the \verb|wimp_capture_accurate| routine is included in the \verb|WimPyC| distribution mainly for validating purposes. However, the accuracy of the \verb|wimp_capture| routine relies on the energy sampling of the response functions, and deserves a few comments. In particular, when called with default input values the \verb|load_response_functions_capture| routine is optimized for a celestial body not much different from the Sun. In several situations the sampling may need to be increased or extended. For instance this may be needed:

\begin{itemize}
    \item close to the boundaries where capture is kinematically allowed, such as in the case of a weak gravitational potential (as in the Earth, for which resonant capture when $m_\chi\simeq m_T$ is observed~\cite{capture_gould_1987}) or for endothermal inelastic scattering ($\delta>0$), when the WIMP speed barely crosses the threshold to excite the heavy state;
    \item a strong gravitational potential: in this case the energy range where the response functions need to be sampled becomes very large, spanning several orders of magnitude (as in the case of a White Dwarf, where they extend to energies much larger than those for the Sun);
    \item a steep energy dependence of the response functions in specific energy ranges (as in the case of a massless propagator at low energies);
    \item a large destructive interference among different interaction operators.
\end{itemize}

Other numerical issues may emerge close to specific regimes where the validity of the effective theory breaks down, such as at small momentum transfers for a massless mediator, where the scattering cross section diverges~\cite{halo_independent_sogang_long_range} and the integrated response functions of Eq.~(\ref{eq:sigma_tilde}) vanish. In such case, besides applying the "Jupiter cut" as explained in Section~\ref{subsec:massless_propagator}, one needs to avoid sampling the response functions in the nonphysical interval of energies where they rise steeply due the cross section divergence (explicitly, this can lead to very large numerical values for the integrated response functions and induce numerical instabilities in the calculation of the differences $\{f(E_R)\}_{E_1}^{E_2} \equiv f(E_2)-f(E_1)$ in Eq.~(\ref{eq:sigma_tilde}) when $f(E_1),f(E_2)\gg f(E_2)-f(E_1)$\footnote{Note that the integrated response functions are defined up to an arbitrary additive constant, since they only appear in differences.}. 

In particular, one should remind that for a given combination of effective operators \verb|WimPyC| tabulates a single set of response functions for each nuclear target. This implies that a given set of response functions may give very accurate results for some celestial bodies, WIMP mass and mass splitting values, but may have an insufficient energy sampling in other cases. This Section is devoted to showing how to verify that the accuracy of a set of response functions is sufficient, and how to proceed to increase it in case of need. 

Consider the specific example of capture in the Sun from $^{27}Al$ in the case of a spin--dependent interaction with massless propagator:

\begin{Verbatim}[frame=single,xleftmargin=0.5cm,xrightmargin=0.5cm,commandchars=\\\{\}]
model=WD.eft_hamiltonian('c4, qm2',
\{(4, 'qm2'): lambda q, c0,c1: [c0/q**2, c1/q**2]\})
WD.load_response_functions_capture(WD.Al27,model,
n_sampling=100)
\end{Verbatim}

\noindent By default the \verb|load_response_functions_capture| routine assumes \verb|n_samp|
\verb|ling=1000|
and in the example above it is assumed \verb|n_sampling=100| for illustration purposes.
If the file \verb|27Al_c_4_qm2_c_4_qm2.npy| is not already present in the folder \verb|WimPyDD/WimPyC/Response_functions/spin_1_2/| the \verb|load_response_functions_capture| call creates it, tabulating the response functions with 100 energy values and assuming by default a maximal escape velocity inside the celestial body of 2000 km/sec. Using such response functions, a \verb|wimp_capture| call yields for $m_\chi$ = 50 GeV and $\delta$ = 0 the following capture rate in the Sun for a WIMP--proton interaction ($c^0$ = 1, $c^1$ = 1):

\begin{Verbatim}[frame=single,xleftmargin=0.5cm,xrightmargin=0.5cm,commandchars=\\\{\}]
capture=WD.wimp_capture(WD.Sun, model, vmin, delta_eta, 
mchi=50, targets_list=[WD.Al27], v_cut=18.5, c0=1, c1=1)
2.305574123338698e+36
\end{Verbatim}

Is such a result reliable? Of course, the most straightforward way to verify it is to use the \verb|wimp_capture_accurate| routine, that yields:

\begin{Verbatim}[frame=single,xleftmargin=0.5cm,xrightmargin=0.5cm,commandchars=\\\{\}]
WD.wimp_capture_accurate(WD.Sun, model, vmin, delta_eta,
mchi=50, targets_list=[WD.Al27], v_cut=18.5, c0=1, c1=1)
2.2469258764605145e+36
\end{Verbatim}

\noindent i.e.  the two results are comfortingly close within a few percent. Notice that on a MacBook Pro M1 with 16 GB memory the \verb|wimp_capture| call requires less than 10 seconds, while the \verb|wimp_capture_accurate| one about 15 minutes.  Suppose that now one needs to calculate the capture rate on Earth. In such case \verb|WimPyC| will use the response functions calculated above, yielding:

\begin{Verbatim}[frame=single,xleftmargin=0.5cm,xrightmargin=0.5cm,commandchars=\\\{\}]
WD.wimp_capture(WD.Earth, model, vmin, delta_eta,
mchi=50, targets_list=[WD.Al27], c0=1, c1=1)
1.3267198919637261e+33
WD.wimp_capture_accurate(WD.Earth, model, vmin, delta_eta,
mchi=50, targets_list=[WD.Al27], c0=1, c1=1)
1.2986444860763012e+33
\end{Verbatim}

\noindent In this case the two results are still within about 2.2\%.
In such case one can increase the sampling, as described at the end of  Section~\ref{sec:response_functions}:

\begin{Verbatim}[frame=single,xleftmargin=0.5cm,xrightmargin=0.5cm,commandchars=\\\{\}]
WD.wimp_capture(WD.Earth, model, vmin, delta_eta,
mchi=50, targets_list=[WD.Al27], increase_sampling=True, 
c0=1, c1=1)
1.3187031923928676e+33
\end{Verbatim}

\noindent dropping the difference below 2\%.

\begin{figure}[h!]
    \centering
    \includegraphics[width=0.48\columnwidth]{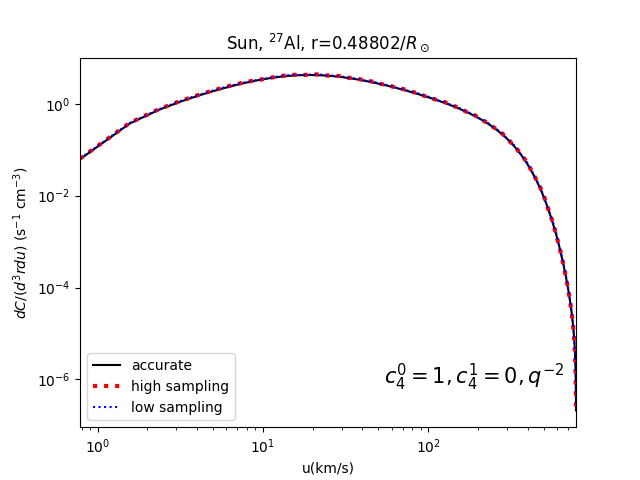}
    \includegraphics[width=0.48\columnwidth]{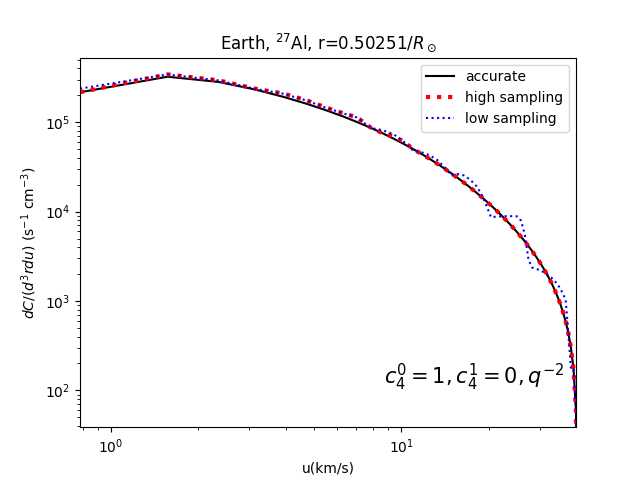}
    \includegraphics[width=0.48\columnwidth]{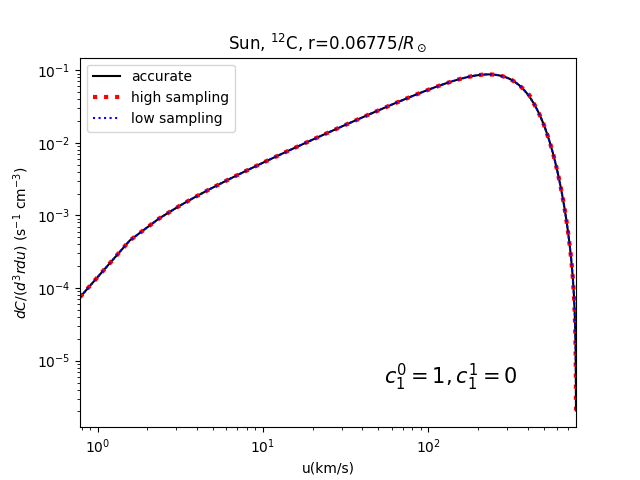}
    \includegraphics[width=0.48\columnwidth]{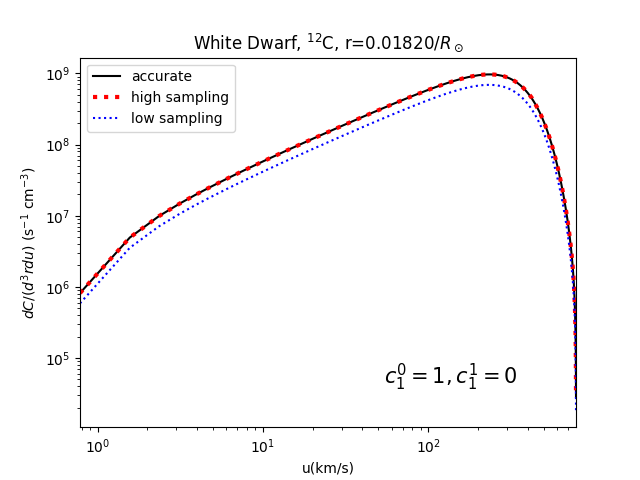}
    \cprotect\caption{{\bf Upper panels:} $\frac{dC^T_{ij}}{d^3 r du}(r,u)$ vs. $u$ at fixed $r$ = 0.5 in units of the celestial body radius for $^{27}Al$ and $m_\chi$ = 50 GeV in the Sun (left) and the Earth (right) and for a SD interaction with a massless mediator. {\bf Lower panels:} Same as upper panels for $^{12}C$ in the Sun (left) and the White Dwarf (right) for a SI contact interaction. }
    \label{fig:dc_dvi}
\end{figure}

Even without using \verb|wimp_capture_accurate|, an indication of under--sampling issues can be provided by a noisy output for the differential capture rate $\frac{dC^T_{ij}}{d^3 r du}(r,u_k)$. The output for the Sun and the Earth in the three cases (accurate calculation, low--sample and high--sample response functions) is shown in the two upper plots of Fig.~\ref{fig:dc_dvi}. One can clearly see that, indeed, while in the case of the Sun a default sampling of 100 points is sufficient to reproduce the differential capture rate accurately, for the Earth it shows some numerical instabilities. In this case, although they do not affect the accuracy in a significant way, one can nevertheless reduce them by improving the sampling with additional 1000 points, following the instructions provided at the end of Section~\ref{sec:response_functions}.

The two plots at the bottom of Fig.~\ref{fig:dc_dvi} show a second example, corresponding to a contact interaction driven by the ${\cal O}_1$ operator in the Sun and in a White Dwarf. In these two plots the response functions are calculated following the same procedure of the upper plots. In particular, one can see that, as far as the Sun is concerned, a simple evaluation of the response functions with default input parameters (low sampling) is sufficient to calculate the signal accurately (the evaluations of the capture rate in the three cases, accurate, low sampling and high sampling, differ at the $10^{-4}$ level). However, in the case of the White Dwarf the low-sampling result underestimates the signal by about 30\%. In this case the problem is that in a White Dwarf the escape velocity is much larger than the default value of 2000 km/s assumed in \verb|load_response_functions_capture| for the maximal escape velocity, in particular implying recoil energies that extend beyond those needed for the Sun. The reason why a larger value is not assumed by default by \verb|load_response_functions_capture| is to avoid a too large energy interval, for which the default sampling would be too sparse. As shown in the plot, the high--sampling evaluation, obtained by calling \verb|wimp_capture| with \verb|increase_sampling=True| is sufficient to fix the problem: the capture rate in the accurate and high sampling results differ, again, at the $10^{-4}$ level.

\end{document}